\documentclass[aps,reprint,superscriptaddress,floatfix,longbibliography]{revtex4-2}

\usepackage{slashed}
 \usepackage{amsmath}
\usepackage{mathtools}
\usepackage{amssymb}
\usepackage{graphicx,bm}
\usepackage{color}
\usepackage[squaren]{SIunits}
\usepackage{microtype}
\usepackage[colorinlistoftodos, color=Green]{todonotes}

\usepackage{ulem}

\usepackage[dvipsnames]{xcolor}

\usepackage{verbatim}

\usepackage{grffile}

\graphicspath{{./figures/}}

\newcommand{\Lagr}{\mathcal{L}}

\usepackage{hyperref}
\hypersetup{
    colorlinks=true,
    linkcolor=blue,
    citecolor=red,
    filecolor=magenta,      
    urlcolor=cyan,
}

\begin{document}

\title{Back-reflection in dipole fields and beyond}

\author{Maksim Valialshchikov}
\email[]{maksim.valialshchikov@uni-jena.de}
\affiliation{Helmholtz Institute Jena, Fröbelstieg 3, 07743 Jena, Germany}
\affiliation{Institute of Optics and Quantum Electronics, Friedrich-Schiller-Universität, Max-Wien-Platz 1, 07743 Jena, Germany}

\author{Felix Karbstein}
\email[]{felix.karbstein@uni-jena.de}
\affiliation{Helmholtz Institute Jena, Fröbelstieg 3, 07743 Jena, Germany}
\affiliation{GSI Helmholtzzentrum für Schwerionenforschung GmbH, Planckstrasse 1, 64291 Darmstadt, Germany}
\affiliation{Theoretisch-Physikalisches Institut, Abbe Center of Photonics, Friedrich-Schiller-Universit\"at Jena, Max-Wien-Platz 1, 07743 Jena, Germany}

\author{Daniel Seipt}
\email[]{d.seipt@hi-jena.gsi.de}
\affiliation{Helmholtz Institute Jena, Fröbelstieg 3, 07743 Jena, Germany}
\affiliation{GSI Helmholtzzentrum für Schwerionenforschung GmbH, Planckstrasse 1, 64291 Darmstadt, Germany}
\affiliation{Institute of Optics and Quantum Electronics, Friedrich-Schiller-Universität, Max-Wien-Platz 1, 07743 Jena, Germany}

\author{Matt Zepf}
\affiliation{Helmholtz Institute Jena, Fröbelstieg 3, 07743 Jena, Germany}
\affiliation{GSI Helmholtzzentrum für Schwerionenforschung GmbH, Planckstrasse 1, 64291 Darmstadt, Germany}
\affiliation{Institute of Optics and Quantum Electronics, Friedrich-Schiller-Universität, Max-Wien-Platz 1, 07743 Jena, Germany}

\date{\today}

\begin{abstract}

Quantum reflection is a fascinating signature of the quantum vacuum that emerges from inhomogeneities in the electromagnetic fields. In pursuit of the prospective real-world implementation of quantum reflection in the back-reflection channel, we provide the first numerical estimates for the light-by-light scattering with dipole pulses, which are known to provide the tightest focusing of light possible. For an all-optical setup with a dipole pump and Gaussian probe of the same frequency, we find that the dominant signal signature is related mainly to the back-reflection channel from 4-wave mixing.
Focusing on this, we study the particular case of a multiple focusing pulses configuration (belt configuration) as an approximation to the idealized dipole pulse. Using Bayesian optimization methods, we determine optimal parameters that maximize the detectability of a discernible back-reflection signal.
Our study indicates that the optimization favors a three-beam collision setup, which we further investigate both numerically and analytically.

\end{abstract}

\maketitle


\section{Introduction}
\label{sec:intro}

Quantum vacuum fluctuations give rise to effective nonlinear interactions between macroscopic electromagnetic fields. Within the Standard Model of particle physics, quantum electrodynamics (QED) describes the leading quantum vacuum nonlinearities originating from the coupling of electromagnetic fields via a virtual electron-positron loop \cite{fedotov2023advances,marklund2006nonlinear,king2016measuring,karbstein2020probing}. This interaction may alter the polarization, frequency, and wave vector of the incident photons through the fundamental quantum process known as ``photon-photon scattering'' or ``light-by-light scattering'' \cite{Karplus:1950zza,Karplus:1950zz, DeTollis:1964una}.

At intensities far below the Schwinger limit, these effects are typically very small (signals on the few-photon level), especially compared to the huge laser background. Therefore, measuring light-by-light scattering with real photons is a significant experimental challenge. However, the development of modern laser and detector technologies makes it possible to propose \cite{King:NatPhot2010,king2012photon,shen2018exploring,karbstein2018vacuum,wang_exploring_2024,Ahmadiniaz:2024xob} and perform \cite{Moulin:1996vv,Moulin:1999hwj,Bernard:2000ovj, Watt:2024brh, Ahmadiniaz:2024xob} the first potentially successful light-by-light scattering experiments with \textit{real on-shell} photons; for experiments studying light-by-light scattering phenomena using strong Coulomb fields see \cite{Leo:1975fb, jarlskog1973measurement, Akhmadaliev, CMS:2018erd, ATLAS:2019azn}.

Since the signals are typically very small, one is interested in finding discernible signatures that differ from the background in their polarization \cite{Karbstein:2019bhp}, angular \cite{Tommasini:2010fb, King:NatPhot2010, Karbstein:2020gzg, Macleod:2024jxl} or frequency \cite{lundin2006analysis, King:2018wtn, Gies:2021ymf, Berezin:2024fxt} properties. A particularly interesting discernible signature is produced via quantum reflection, which was first suggested by \cite{Gies:2013yxa, Gies:2014wsa}. The authors found that when a probe photon collides with an electromagnetic field inhomogeneity of sufficient intensity and localization, there is a nonzero probability for the probe photon to be reflected. They also provided the first analytic estimates of the magnitude of the effect, but, of course, these were limited by simplifying assumptions regarding the electromagnetic field profile, e.g., the use of an infinite Rayleigh range approximation \cite{Gies:2017ygp}.

To study more realistic field configurations and provide more accurate estimates, one can use numerical Maxwell solvers to model the background fields. For instance, the authors of \cite{grismayer2021quantum, lindner2023numerical, zhang2025computational} use nonlinear Maxwell solvers to model the nonlinear interaction of the fields. We resort to the numerical approach first implemented in \cite{blinne2019all} that uses a linear Maxwell solver to model the background fields and the \textit{vacuum emission picture} \cite{karbstein2015stimulated} to calculate the quantum vacuum signal. 
In the \textit{vacuum emission picture}, the signal is separated from the large background, making it a convenient calculation tool that was applied in various scenarios \cite{fedotov2023advances} and could be coupled in a straightforward way with numerical optimization \cite{Valialshchikov:2024svm}.

In this article, we look for the prospective real-world implementation of quantum reflection that could yield a sizable signal. 
Dipole pulses \cite{gonoskov2012dipole, jeong20204pi} have high focusing efficiency and have been argued to benefit $e^+ e^-$ pair creation \cite{Gonoskov:2013aoa, Gonoskov:2016ynx}. They have also been considered for high-order harmonic generation \cite{Sasorov:2025jjr} and light-induced gravitational effects \cite{Fillion-Gourdeau:2025eml}. Motivated by this, we consider a dipole pulse to be a promising candidate for the back-reflection of incoming probe photons. We provide the first numerical estimates for the light-by-light scattering with dipole pulses and identify several configurations yielding the largest reflected signal. For the equal frequency case, we find that this reflected signal mainly originates from the back-reflected channel of 4-wave mixing and focus on this signature in more detail. For this case, due to the $4 \pi$-focusing of a dipole pulse, the signals remain background-dominated. To address this issue, we study a multiple colliding pulses configuration as a rough approximation to the dipole fields. Such a scenario indeed makes the back-reflected signal discernible, albeit at the expense of its magnitude. Additionally, we provide optimal parameters for the probe orientation and polarization of belt pulses to maximize the discernible back-reflected signal. Optimizing energy distribution between belt pulses shows that an even larger signal could be achieved with just three pulses. Focusing on a subset of three pulse collisions originating from our belt setup, we provide numerical and analytical estimates for this case.

Our paper is structured as follows. We briefly summarize the vacuum emission picture in Section \ref{sec:vep}, dipole fields in Section \ref{sec:dipole-theory}, and details about the setting we consider in Section \ref{sec:setting}. Numerical simulations are described in Section \ref{sec:numerics:simulation} and details about Bayesian optimization -- in Section \ref{sec:numerics:optimization}. We begin the results section by studying the collision of a Gaussian pulse with a dipole pulse in Section \ref{sec:dipole}. This is followed by considering a belt configuration in Section \ref{sec:belt} and a three pulse collision in Section \ref{sec:3-beams}. Finally, we end with conclusions in Section \ref{sec:conclusions}.

Aside from a few exceptions where we intend to highlight the explicit dependence on $c$ and $\hbar$, throughout this work we use the Heaviside-Lorentz system with natural units $\hbar=c=\epsilon_0=1$.

\section{Formalism}
\label{sec:formalism}

\subsection{Vacuum emission picture}
\label{sec:vep}

Effective nonlinear interactions of electromagnetic fields are described by the Heisenberg-Euler Lagrangian ${\cal L}_{\rm HE}$ \cite{Heisenberg:1936nmg}. For weak fields satisfying $\{|\mathbf{E}|,c|\mathbf{B}|\} \ll E_{\rm cr}$, with critical electric field strength $E_{\rm cr} = m_e^2 c^3 / (e \hbar) \simeq 1.3 \times 10^{18}\,{\rm V/m}$ and varying on typical spatiotemporal scales $\lambda \gg \lambdabar_{\rm C}$ with electron Compton wavelength $\lambdabar_{\rm C}=\hbar/(m_ec)\simeq3.8\times10^{-13}\,{\rm m}$ the leading contribution is given by \cite{euler1935scattering}
\begin{equation}
    \Lagr_{\text{HE}}^{1\text{-loop}} \simeq \frac{m_e^4}{360 \pi^2} \left(\frac{e}{m_e^2}\right)^4 (4 \mathcal{F}^2 + 7 \mathcal{G}^2)\,,
    \label{eq:HE-lagrangian}
\end{equation}
where $e$ is the elementary charge, $m_e$ is the electron mass and $\mathcal{F}=\frac{1}{4} F_{\mu \nu} F^{\mu \nu} = \frac{1}{2} (\mathbf{B}^2 - \mathbf{E}^2)$, $\mathcal{G}=\frac{1}{4} F_{\mu \nu} \prescript{\star}{ }{F^{\mu \nu}}=-(\mathbf{B} \cdot \mathbf{E})$ are the electromagnetic field invariants. The leading-order approximation is well justified for electromagnetic fields attainable in current and near-future optical laser experiments.

The vacuum emission picture \cite{karbstein2015stimulated} is a convenient way to calculate quantum vacuum signals. The zero-to-single signal photon transition amplitude to a state characterized by a wave vector $k^\mu=(\omega,{\bf k})$, with $\omega=|{\bf k}|$, and a polarization vector $\epsilon_{(p)}^{\mu}(k)$ can be expressed as
\begin{equation}
    S_{(p)}(\mathbf{k}) = \frac{\epsilon_{(p)}^{*\mu}(k)}{\sqrt{2 k^0}} \int d^4 x\, e^{ik_{\alpha} x^{\alpha}} j_{\mu}(x)\bigg|_{k^0=\omega},
    \label{eq:amplitude}
\end{equation}
where
\begin{equation}
    j_{\mu}(x) = 2 \partial^{\nu} \frac{\partial \Lagr_{\text{HE}}}{\partial F^{\nu \mu}}
    \label{eq:vacuum-current}
\end{equation}
is the signal-photon current induced by the applied macroscopic electromagnetic fields $F^{\mu \nu}$.

Substituting Eq. \eqref{eq:HE-lagrangian} to Eqs. \eqref{eq:amplitude} and \eqref{eq:vacuum-current}, we obtain for the signal amplitude \cite{Gies:2021ymf}
\begin{align}
    S_{(p)}(\mathbf{k}) =& \: i \frac{e}{4 \pi^2} \frac{m_e^2}{45} \left( \frac{e}{m_e^2} \right)^3 \frac{\epsilon_{(p)}^{*\mu}(k)}{\sqrt{2 k^0}} \int d^4 x\, e^{ik_{\alpha} x^{\alpha}} \nonumber \\
    &\times \left( 4 k^{\nu} F_{\nu\mu} \mathcal{F} + 7 k^{\nu} \prescript{\star}{ }{F_{\nu \mu} \mathcal{G}} \right)\bigg|_{k^0=\omega},
\end{align}
which has a cubic dependence on the electromagnetic field strength tensor. 

For several colliding pulses, it is often useful for the analysis to split the signal amplitude into different channel contributions. For later reference, we already want to introduce this notation here. Omitting Lorentz indices, the signal amplitude can be schematically written as
\begin{equation}
    \label{eq:channel-notation}
    S_{(p)}(\mathbf{k}) = \sum_{i,j,k} S_{(p)}^{ijk}(\mathbf{k}) \sim \sum_{i,j,k}\int d^4 x\, e^{ik_{\alpha} x^{\alpha}} F_i F_j F_k,
\end{equation}
where $F_i$ corresponds to the electromagnetic field strength tensor of the $i$-th pulse and the sum is performed over all possible combinations.

Modulus squaring the transition amplitude (Eq. \eqref{eq:amplitude}) gives the differential number of signal photons
\begin{equation}
    d^3 N_{(p)}(\mathbf{k}) = \frac{d^3 \mathbf k}{(2\pi)^3} |S_{(p)}(\mathbf{k})|^2.
\end{equation}

The polarization-insensitive signal photon spectrum is obtained by summing over the transverse polarizations: $d^3 N(\mathbf{k}) = \sum_p d^3 N_{(p)}(\mathbf{k})$. In this article, we focus only on polarization-insensitive signals (for which we use the notation $S^{ijk}$ similar to Eq. \eqref{eq:channel-notation}). The integration over all possible signal-photon energies gives the polarization-insensitive angular resolved signal photon density
\begin{equation}
    \frac{d^2 N}{d^2\Omega}(\vartheta, \varphi) = \int_0^\infty \frac{d\omega \, \omega^2}{(2\pi)^3} \: \sum_p |S_{(p)}(\mathbf{k})|^2\,,
\end{equation}
with solid angle element $d^2\Omega=d\cos\vartheta\,d\varphi$. 

We introduce the analogous quantity for background fields $d^2 N^{\rm bgr}/d^2\Omega(\vartheta, \varphi)$ and compare it with the signal one to determine \textit{angularly discernible} regions:
\begin{equation}
    \frac{d^2 N}{d^2\Omega}(\vartheta, \varphi) > \frac{d^2 N^{\text{bgr}}}{d^2\Omega}(\vartheta, \varphi).
\end{equation}

Integration over these regions yields the total discernible signal -- $N_{\rm disc}$. Specifying the angular region of interest by $\Omega_{\rm det}$ and integrating over it gives us the total ``detected'' signal
\begin{equation}
    N_{\rm det} = \int_{\Omega_{\rm det}} d \Omega \frac{d^2 N}{d^2\Omega}(\vartheta, \varphi).
\end{equation}


\subsection{Dipole fields}
\label{sec:dipole-theory}

Dipole pulses are exact and singularity-free solutions of the free Maxwell equations, providing efficient focusing of electromagnetic field energy \cite{gonoskov2012dipole, jeong20204pi}. Compared to focused Gaussian pulses, optimal focusing of dipole pulses leads to higher field amplitudes, which can result in the enhancement of light-by-light scattering signatures \cite{Ilderton:2016khs}. Strong field localization can be especially useful for enhancing quantum reflection.

A dipole pulse is characterized by its type (electric dipole $\leftrightarrow$ $e$-dipole or magnetic dipole $\leftrightarrow$ $b$-dipole), the orientation of its virtual dipole moment $\mathbf{d}$, its wavelength, and its duration. In an $e$-dipole, the electric field dominates at the focus; in a $b$-dipole, the magnetic field does. 

The electromagnetic field of an $e$-dipole pulse is given by (Eqs. (23a) and (23b) from \cite{gonoskov2012dipole}):
\begin{subequations}
\label{eq:dipole-EB}
\begin{align}
    \mathbf{B} (t, \mathbf{r}) =& -[\mathbf{n} \times \mathbf{d}] \left[ \frac{1}{c^2} \frac{\ddot{g}_+(t,r)}{r} + \frac{1}{c} \frac{\dot{g}_-(t,r)}{r^2} \right], \\
    \mathbf{E} (t, \mathbf{r}) =& \frac{\mathbf{n} \times [\mathbf{n} \times \mathbf{d}]}{r c^2} \ddot{g}_-(t,r) + \frac{3 \mathbf{n} (\mathbf{n} \cdot \mathbf{d}) - \mathbf{d}}{r^3} \\
    & \times \left[ \frac{r}{c} \dot{g}_+(t, r) + g_-(t,r) \right], \nonumber
\end{align}
\end{subequations}
where $r = |\mathbf{r}|$, $\mathbf{n} = \mathbf{r} / r$ is the unit radius vector, $g_{\pm}(t,r) = g(t - r/c) \pm g(t + r/c)$ and $g(t)$ is an arbitrary function. In this work, we consider only dipole pulses with a Gaussian temporal envelope $g(t) = \exp(-t^2/(\tau/2)^2) \exp(-i \omega t)$, where $\tau$ is the dipole pulse duration measured at $1/e^2$ of the intensity and $\omega$ is its frequency.

The field distribution of a magnetic dipole pulse can be obtained from Eq. \eqref{eq:dipole-EB} by using the following rule \cite{gonoskov2012dipole}:
\begin{equation}
    \mathbf{E}^{e-{\rm dipole}} \rightarrow \mathbf{B}^{b-{\rm dipole}},\:\: \mathbf{B}^{e-{\rm dipole}} \rightarrow -\mathbf{E}^{b-{\rm dipole}}.
\end{equation}

\begin{figure}
    \centering
    \includegraphics[width=1\linewidth]{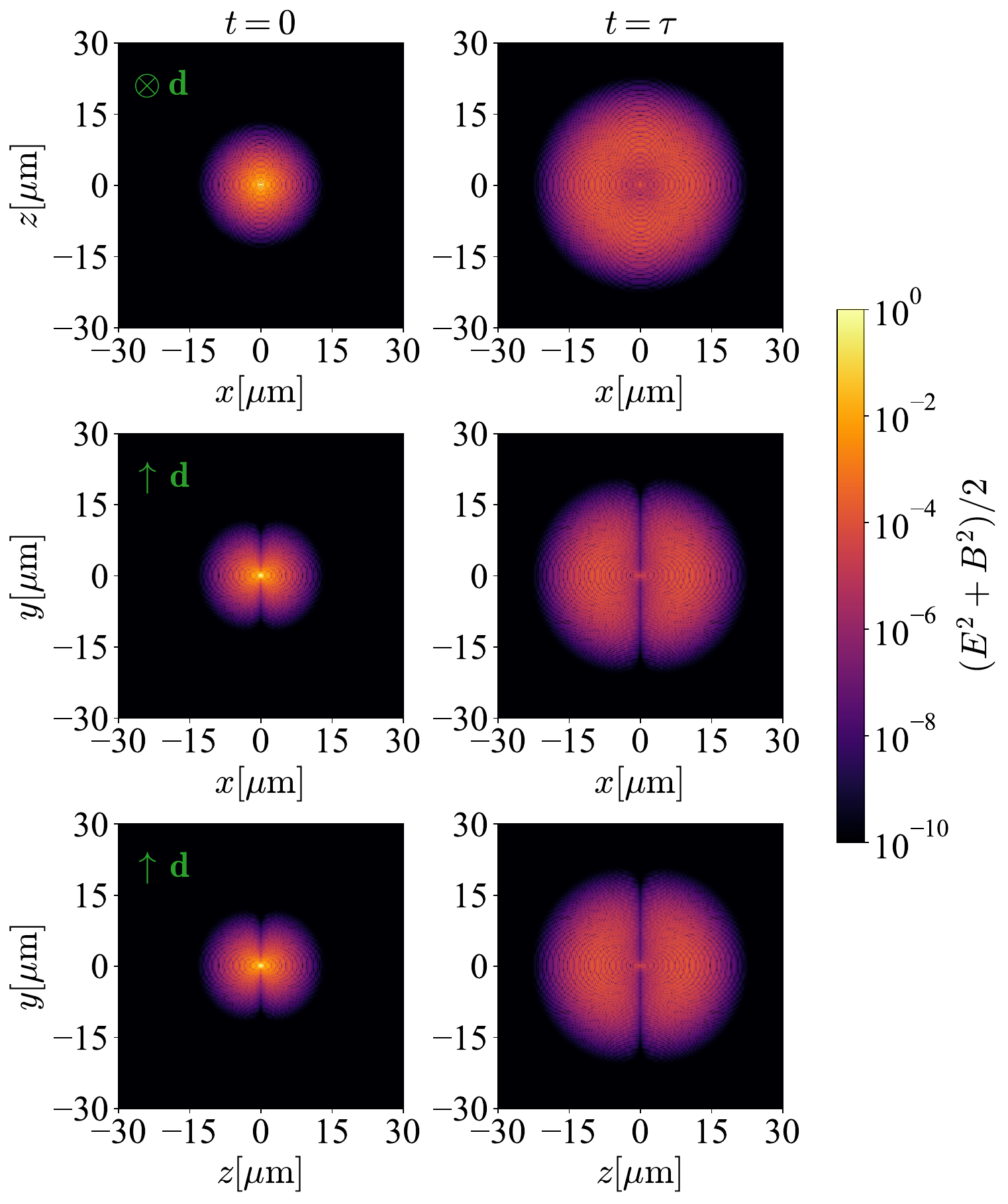}
    \caption{Electromagnetic energy density of a magnetic dipole pulse in three planes ($xz$ at $y=0$, $xy$ at $z=0$, and $yz$ at $x=0$) at two time steps: (left column) $t = 0$ (focus), (right column) $t = \tau$. Dipole pulse parameters: $W = 40\, \rm{J}$, $\lambda = 800\, \rm{nm}$, $\mathbf{d} \: \| \: \mathbf{e}_y$, $\tau_{\rm{FWHM}} = 20\, \rm{fs}$.
    }
    \label{fig:dipole-focus}
\end{figure}

Figure \ref{fig:dipole-focus} shows the dipole pulse's electromagnetic energy density, which is radially symmetric in the plane perpendicular to the dipole moment (xz) and forms a characteristic dipole emission pattern in other planes (xy and yz). The effective focusing volume of energy density concentration of a dipole pulse is much smaller than $\lambda^3$ for focused Gaussian pulses, therefore, achieving higher field amplitudes and localization \cite{gonoskov2012dipole}. For our parameters, the rough estimate of the volume bounded by the half height of the energy density concentration is $V_{\rm{dp}} = l_{\perp}^2 l_{||} \approx 0.093 \lambda^3$, where  $l_{\perp} \approx 0.4 \lambda$ and $l_{||} \approx 0.6 \lambda$ are the characteristic extensions in the transverse and longitudinal directions respectively (FWHM). Such effective focusing of electromagnetic field energy was our main motivation to consider a dipole pulse as a promising candidate for quantum reflection.

\subsection{Considered setting}
\label{sec:setting}

We are interested in the optical (and near-IR) frequency range with all pulses originating from PW-class lasers and colliding at zero impact parameter. 
In this article, we consider pulses with energies ranging from $W = 10\, \rm{J}$ to $W = 40\, \rm{J}$, the wavelength $\lambda = 800\, \rm{nm}$ and the pulse duration $\tau_{\rm FWHM} = 20\, \rm{fs}$ measured at full width half maximum (FWHM) of the intensity. In the simulations, we use a Gaussian temporal profile $\exp(-t^2/(\tau/2)^2)$, where $\tau$ is the pulse duration measured at $1/e^2$ of the intensity. These durations are related by $\tau = \sqrt{2/\ln 2}\: \tau_{\rm FWHM}$. The polarization of Gaussian pulses is determined by the angle $\beta$, corresponding to $\sphericalangle \{\mathbf{E}(t=0,\mathbf{r}=\mathbf{0}), \mathbf{e}_x\}$ in the frame of reference where the pulse propagates along $\mathbf{e}_z$. Here $\mathbf{e}_x$ and $\mathbf{e}_z$ are the basis vectors for $x$ and $z$ axis respectively.

In all considered scenarios, we provide the total vacuum emission signal and name one pulse ``the probe'' only for notational convenience. Let us note that we do not perform any linearization in the probe field at the amplitude level at this stage because for some setups there are several channels contributing to the spectral region of interest. As an observable, we focus on the back-reflected signal with particular interest in the contribution generated by the probe pulse. We determine it with subsequent channel analysis and explicitly state if there are several channels contributing to the back-reflected region.  
To quantify the amount of back-reflected signal, we use a $10^{\circ} \times 10^{\circ}$ detector, which is large enough to capture the relevant region, and study the signal dependence on the parameters of the pulses. At the same time, it seems experimentally reasonable: for instance, a $10\,\rm{mm}$ camera chip located $1\,\rm{m}$ from the interaction point would require a $17\times$ magnification system.

\section{Numerical methods}
\label{sec:numerics}

\subsection{Simulation}
\label{sec:numerics:simulation}

The algorithm to obtain realistic quantum vacuum signal estimates was originally put forward in \cite{blinne2019all}. Their code employed 1) a linear Maxwell solver to describe the evolution of the external electromagnetic fields and 2) the vacuum emission picture to determine the leading quantum vacuum signal via Eq.~\eqref{eq:amplitude}. Such an approach does not rely on simplifying assumptions and approximations regarding the spatiotemporal structure of the participating fields and can provide quantitatively accurate results in experimentally relevant configurations. 

To extend the physics case and provide a range of additional utilities, we wrote our own code \textit{quvac} \cite{quvac} that relies on the same core algorithm: linear Maxwell solver + vacuum emission picture. To remove some limitations of the previous code, we added the possibility to calculate channel-separated signals and use any combination of analytic and Maxwell-propagated fields. Additional modularity and postprocessing utilities allowed us to achieve almost seamless integration of grid scans and Bayesian optimization. For faster computations, we added the possibility of lower precision calculations (float32) \cite{Gies:2024aaz} and parallelization.

We initialize the self-consistent Maxwell propagation by constructing manifestly transverse fields based on an initial model field configuration at focus ($t=0$) in the space domain (for details, see Section III D of \cite{blinne2019all}). This field is then propagated to other time steps according to the linear Maxwell equations. For Gaussian pulses, we use the leading-order paraxial Gaussian \cite{salamin2007fields} as the model field in the focus; for dipole pulses, we use Eq. \eqref{eq:dipole-EB} with a Gaussian temporal envelope.

\subsection{Bayesian optimization}
\label{sec:numerics:optimization}

Following \cite{Valialshchikov:2024svm}, we use Bayesian optimization to efficiently explore the parameter landscape of configurations with several free parameters. We use the optimization framework ``Ax'' \cite{ax} which provides a high-level implementation of Bayesian optimization. This framework is integrated into our \textit{quvac} package.

We start with 5 to 15 randomly sampled observations, serving as a foundation for the Bayesian model initialization, and continue up to several tens of Bayes steps. The optimization is dynamically parallelized between several computational nodes. 

\section{Results}

\subsection{Dipole pulse}
\label{sec:dipole}

As described in Section \ref{sec:dipole-theory}, a dipole pulse has a very high focusing efficiency of electromagnetic field energy and, as a result, high field amplitudes and high intensity gradients around the focus. Such field properties make it a promising candidate for quantum reflection. In this section, we study the characteristics and estimate the magnitude of the reflected signal in the collision of a Gaussian probe pulse and a dipole pump pulse.

In this section, we consider a Gaussian probe pulse with energy $W = 20\,\rm{J}$, wavelength $\lambda = 800\,\rm{nm}$, duration $\tau_{\rm FWHM}=20\,\rm{fs}$ and waist size $w_0 = 2\lambda$, achieving maximal intensity of $I_{\rm max} = \:u_{\rm max}/2 \approx 2.3 \times 10^{22}\, \rm{W/cm^2}$, where $u=(E^2+B^2)/2$ is the electromagnetic energy density. A magnetic dipole pump pulse has the following parameters: $W = 40\,\rm{J}$, $\lambda = 800\,\rm{nm}$, $\tau_{\rm FWHM}=20\,\rm{fs}$, achieving maximal intensity of $I_{\rm max} \approx 1.2 \times 10^{24}\, \rm{W/cm^2}$. Note that the probe and the pump have the same frequency.

\begin{figure}[t!]
    \centering
    \includegraphics[width=0.95\linewidth]{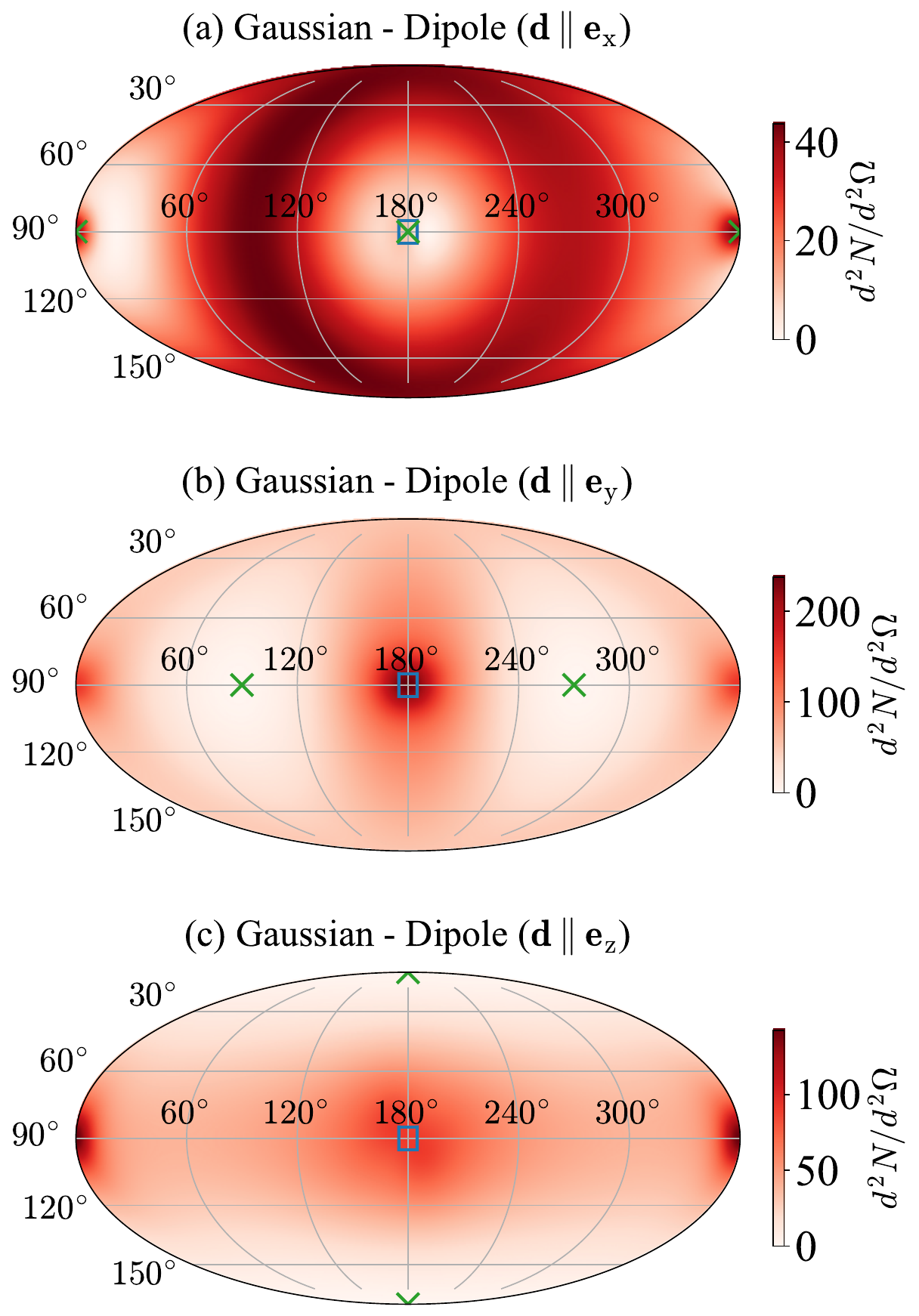}
    \caption{Angular signal photon spectrum from the collision of a Gaussian probe with a dipole pump. Probe pulse has the following parameters: $\hat{\mathbf{k}} = \mathbf{e}_{\rm{x}}$, $W = 20\, \rm{J}$, $\lambda = 800\, \rm{nm}$, $\tau_{\rm{FWHM}} = 20\, \rm{fs}$, $w_0 = 2\lambda$, $\beta = 0^{\circ}$. The probe collides with the magnetic dipole pulse ($W = 40\, \rm{J}$, $\lambda = 800\, \rm{nm}$, $\tau_{\rm{FWHM}} = 20\, \rm{fs}$) with different dipole moment orientations. Green crosses show the dipole moment axis. The blue square shows a $10^{\circ} \times 10^{\circ}$ detector for which we calculate the reflected signal ($N_{\rm{det}}$). (a) $\mathbf{d} \: \| \: \mathbf{e}_x$, $N_{\rm{tot}} \approx 348, \: N_{\rm{det}} \approx 0.25$; (b) $\mathbf{d} \: \| \: \mathbf{e}_y$, $N_{\rm{tot}} \approx 574, \: N_{\rm{det}} \approx 7.2$; (c) $\mathbf{d} \: \| \: \mathbf{e}_z$, $N_{\rm{tot}} \approx 416, \: N_{\rm{det}} \approx 2.8$. 
    }
    \label{fig:gg-gd}
\end{figure}

Figure \ref{fig:gg-gd} shows the angular signal photon spectra for different orientations of the dipole moment. Recall, that when two Gaussian pulses collide, the signal is centered around their directions of propagation.
For the dipole case, we can see that the angular signal spectrum is very broad, and it heavily depends on the dipole moment orientation. Extended ring-like structures around the dipole moment axis (bright for case (a) and dimmer for cases (b) and (c)) originate from the dipole self-scattering (see discussion below).
The forward probe channel around $(\vartheta, \varphi) = (90^{\circ}, 0^{\circ})$ is present for all cases, but the back-reflected signature at $(\vartheta, \varphi) = (90^{\circ}, 180^{\circ})$ is suppressed for some configurations (compare (a) with (b) and (c)).
The nontrivial electromagnetic field structure of the dipole pulse and the interplay between probe polarization and dipole moment orientation lead to different ``preferred'' directions of emission and angular features. For instance, when the linearly-polarized probe ($\beta = 0^{\circ}$) propagates along the dipole moment, this leads to the appearance of an asymmetry with respect to $\varphi$ for case (a). Changing the polarization of the probe to $\beta = 90^{\circ}$ leads to the appearance of an asymmetry with respect to $\vartheta$. Phenomenologically, this asymmetry could have the following origin. For the magnetic dipole, only the magnetic field is present at the focus, and it is aligned with the dipole moment. The electric field lines form rings around the dipole moment axis. For the linearly-polarized probe propagating along the dipole moment, the probe's electric field would be parallel to the dipole's electric field on one side of the ring and anti-parallel on another, leading to the $\varphi$ (or $\vartheta$) asymmetry.

In terms of the total polarization-insensitive signal, the Gaussian-dipole cases (a,b,c) yield $N_{\rm{tot}} \approx (348, 574, 416)$ with $N_{\rm{det}} \approx (0.25, 7.2, 2.8)$ in a $10^{\circ} \times 10^{\circ}$ detector in the back-reflected region of interest.

\begin{figure}
    \centering
    \includegraphics[width=0.95\linewidth]{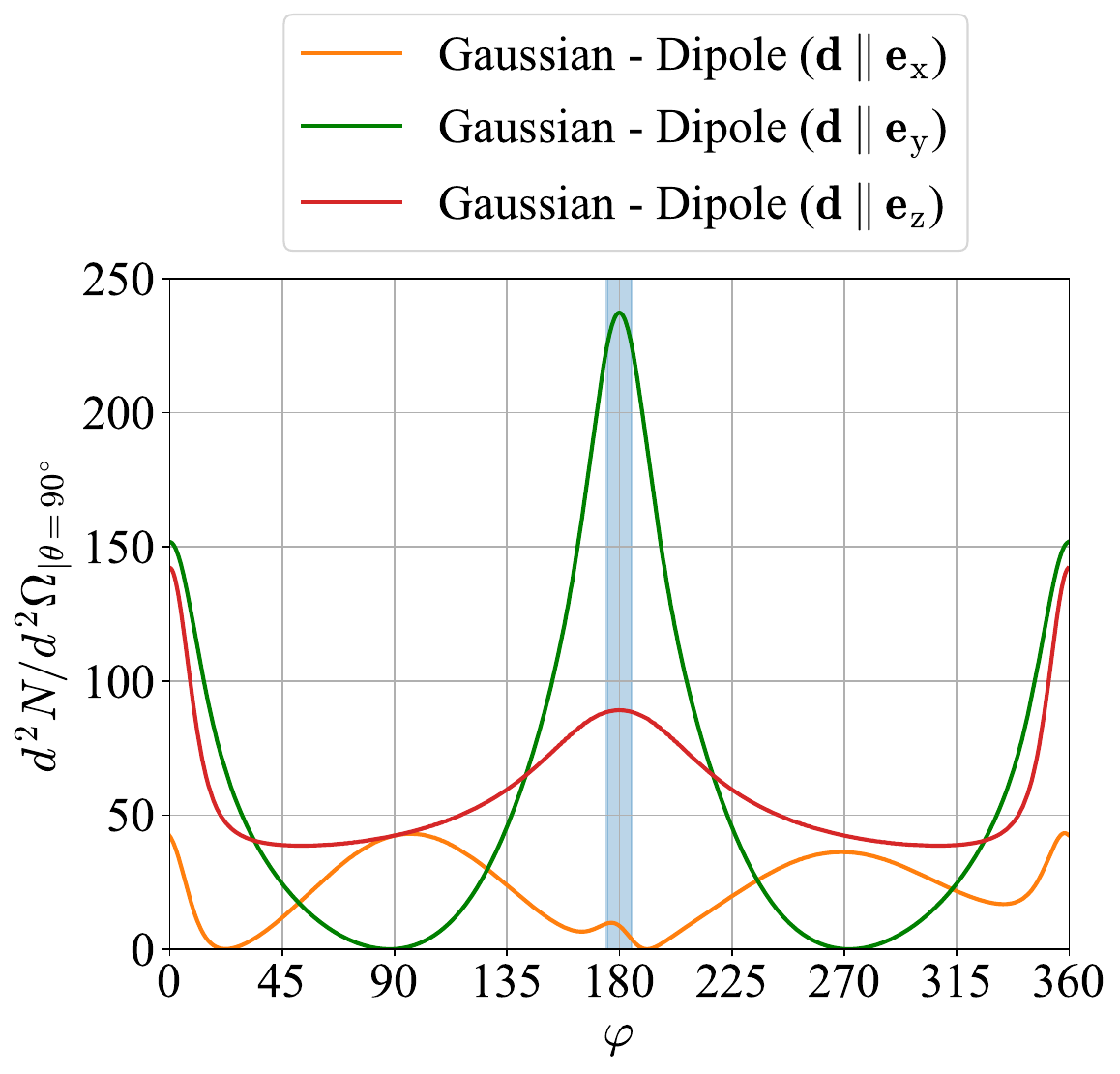}
    \caption{Lineout from the angular signal photon spectrum at $\vartheta = 90^{\circ}$. Different lines correspond to different collision scenarios shown in Fig. \ref{fig:gg-gd}. The blue-shaded region corresponds to the detector for the back-reflected signal.}
    \label{fig:lineout_gg_gd}
\end{figure}

To more clearly see the difference between these cases, Fig. \ref{fig:lineout_gg_gd} shows the lineouts from angular spectra at $\vartheta = 90^{\circ}$. Once again, in all dipole scenarios, the spectrum is very broad. The $\mathbf{d} \: \| \: \mathbf{e}_y$ case has the largest peak in the back-reflected direction, which is even higher than the one in the forward direction (see the channel discussion below for an explanation). 
The left-right asymmetry for the case $\mathbf{d} \: \| \: \mathbf{e}_x$ can be seen here as well.

\begin{figure}
    \centering
    \includegraphics[width=0.9\linewidth]{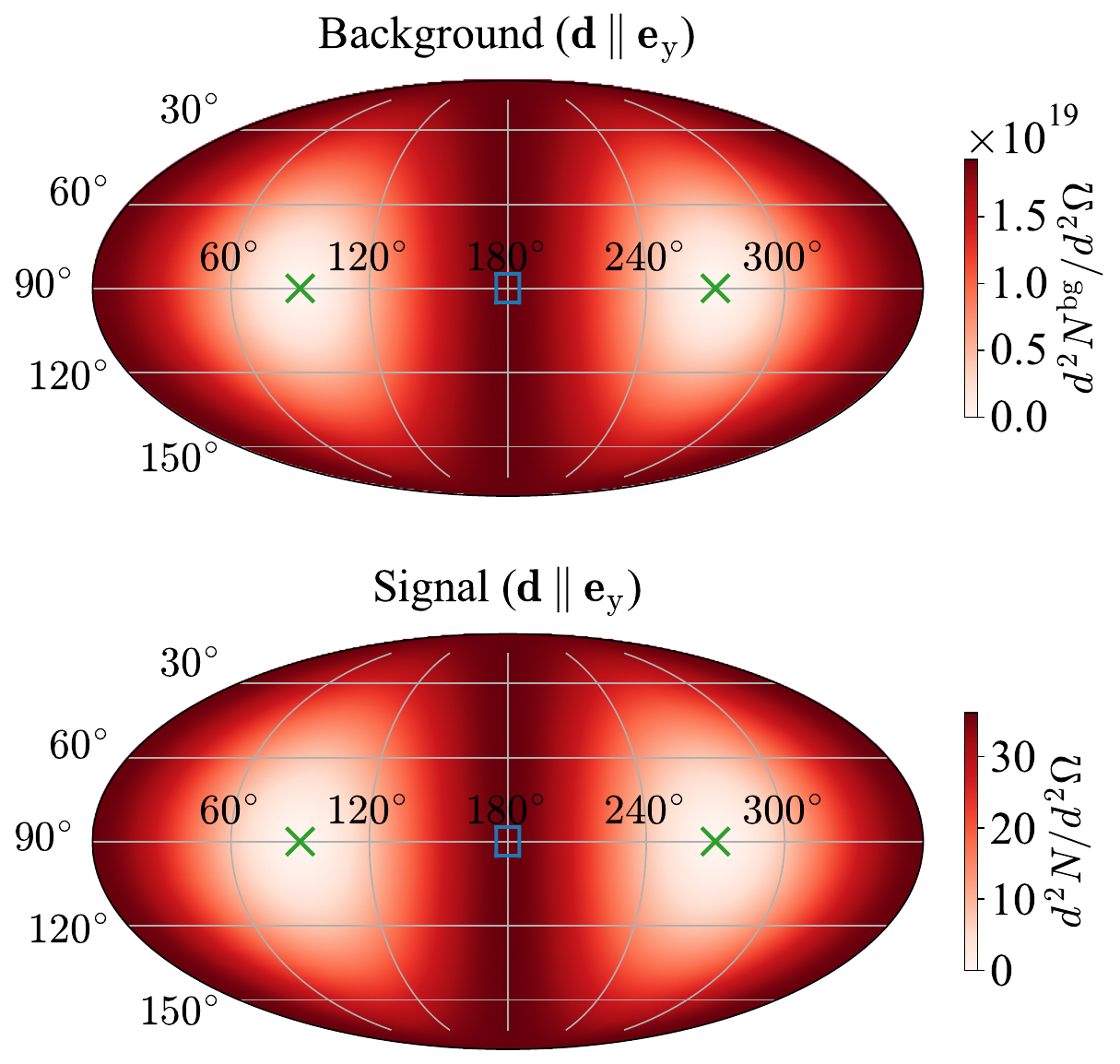}
    \caption{Angular background and signal photon spectrum from a single magnetic dipole pulse. Dipole pulse parameters: $W = 40\, \rm{J}$, $\lambda = 800\, \rm{nm}$, $\mathbf{d} \: \| \: \mathbf{e}_y$, $\tau_{\rm{FWHM}} = 20\, \rm{fs}$. Green crosses show the dipole moment axis. The blue square shows a $10^{\circ} \times 10^{\circ}$ detector. The dipole background results in $N_{\rm{tot}} \approx 1.6 \times 10^{20}$, $N_{\rm{det}} \approx 6.1 \times 10^{17}$ and the dipole self-scattering results in $N_{\rm{tot}} \approx 289$, $N_{\rm{det}} \approx 1.1$.}
    \label{fig:dipole-self-emission}
\end{figure}

Now, let us try to uncover why the angular signal spectrum from the dipole setup has such features. 
Recall that a single paraxial Gaussian pulse has a very small quantum vacuum signal due to self-scattering \cite{blinne2019all}. Conversely, a dipole pulse has a rich $k$-spectrum, resulting in a substantial self-scattering signal, which is shown in Figure \ref{fig:dipole-self-emission} (compare this with Fig. \ref{fig:gg-gd} (b): due to the difference in scales for the signal, dark red in Fig. \ref{fig:dipole-self-emission} corresponds to light orange in Fig. \ref{fig:gg-gd}, but the angular shape of self-scattering remains the same). Similar to the background plot (and the focus planes in Fig. \ref{fig:dipole-focus}), no signal is emitted along the virtual dipole moment, and the slice perpendicular to it exhibits azimuthal symmetry. Note that, due to the $4 \pi$ focusing, the dipole background dominates almost everywhere and our region of interest ($10^{\circ} \times 10^{\circ}$ detector) collected $N_{\rm{det}} \approx 6.1 \times 10^{17}$ photons from the background and $N_{\rm{det}} \approx 1.1$ photons from the self-scattering. For the total signal in the detector $N_{\rm det} \approx 7.2$ from Fig. \ref{fig:gg-gd} (b), this results in the signal-to-background ratio $\sim 1 \times 10^{-17}$. Let us note for completeness that we chose the same frequency for the probe and the dipole in our idealized setup on purpose because such a configuration leads to the largest back-reflected signal due to the energy-momentum conservation (the background issue would be addressed in the next sections). Detuning the frequency of the dipole and probe pulse leads to a rapid decrease in back-reflected signal, showing that this back-reflected signature mainly originates from the 4-wave mixing channel rather than the pure quantum reflection.

Schematically, the dominant channels for the Gaussian-Gaussian (G--G) and Gaussian-dipole (G--D) cases can be written at the amplitude level in the following form (1 stands for the probe and 2 for the pump): $S^{\rm G-G} \approx S^{122} + S^{112}$ ($S^{111}$ and $S^{222}$ are present but highly suppressed),  $S^{\rm G-D} \approx S^{122} + S^{112} + S^{222}$ ($S^{222}$ corresponds to the dipole self-scattering). The interference between these terms plays a substantial role in the signal photon spectrum.
Recall that the $\mathbf{d} \: \| \: \mathbf{e}_y$ case in Figs. \ref{fig:gg-gd} and \ref{fig:lineout_gg_gd} has a larger peak in the back-reflected direction than in the forward one. Numerical channel analysis (for details see Appendix \ref{sec:dipole-channels}) showed that this is due to 1) the $S^{122}$ channel having a larger peak in the back-reflected direction than in the forward one, 2) more favorable interference between $S^{122}, S^{112}$ and $S^{222}$ for the reflected direction, and 3) the $S^{112}$ channel contributing to the reflected direction but not contributing to the forward scattering.

Fixing colliding pulse parameters to the ones from Fig. \ref{fig:gg-gd} and manually varying only the dipole type, the orientation of its dipole moment, and the probe polarization, we found several configurations resulting in the largest back-reflected signal.
The case (b) in Fig. \ref{fig:gg-gd} is one of these configurations: a $b$-dipole pulse with $\mathbf{d} \: \| \: \mathbf{e}_y$ and $\beta_{\rm probe} = 0^{\circ}$ ($\mathbf{E}_{\rm probe} \: || \: -\mathbf{e}_z$, $\mathbf{B}_{\rm probe} \: || \: \mathbf{e}_y$). Let us note that the $\mathbf{d} \: \| \: -\mathbf{e}_y$ case would result in a different signal due to the CEP effect. The same signal is achieved for an $e$-dipole with $\mathbf{d} \: \| \: -\mathbf{e}_z$ and $\beta_{\rm probe} = 0^{\circ}$. Cases with different polarization angles of the probe can be traced down to the aforementioned ones by a rotation of the coordinate frame around the wave vector of the probe: for instance, a $b$-dipole with $\mathbf{d} \: \| \: \mathbf{e}_z$ and $\beta_{\rm probe} = 90^{\circ}$ ($\mathbf{E}_{\rm probe} \: || \: \mathbf{e}_y$, $\mathbf{B}_{\rm probe} \: || \: \mathbf{e}_z$) would result in the same back-reflected signal. 

\begin{figure}
    \centering
    \includegraphics[width=0.75\linewidth]{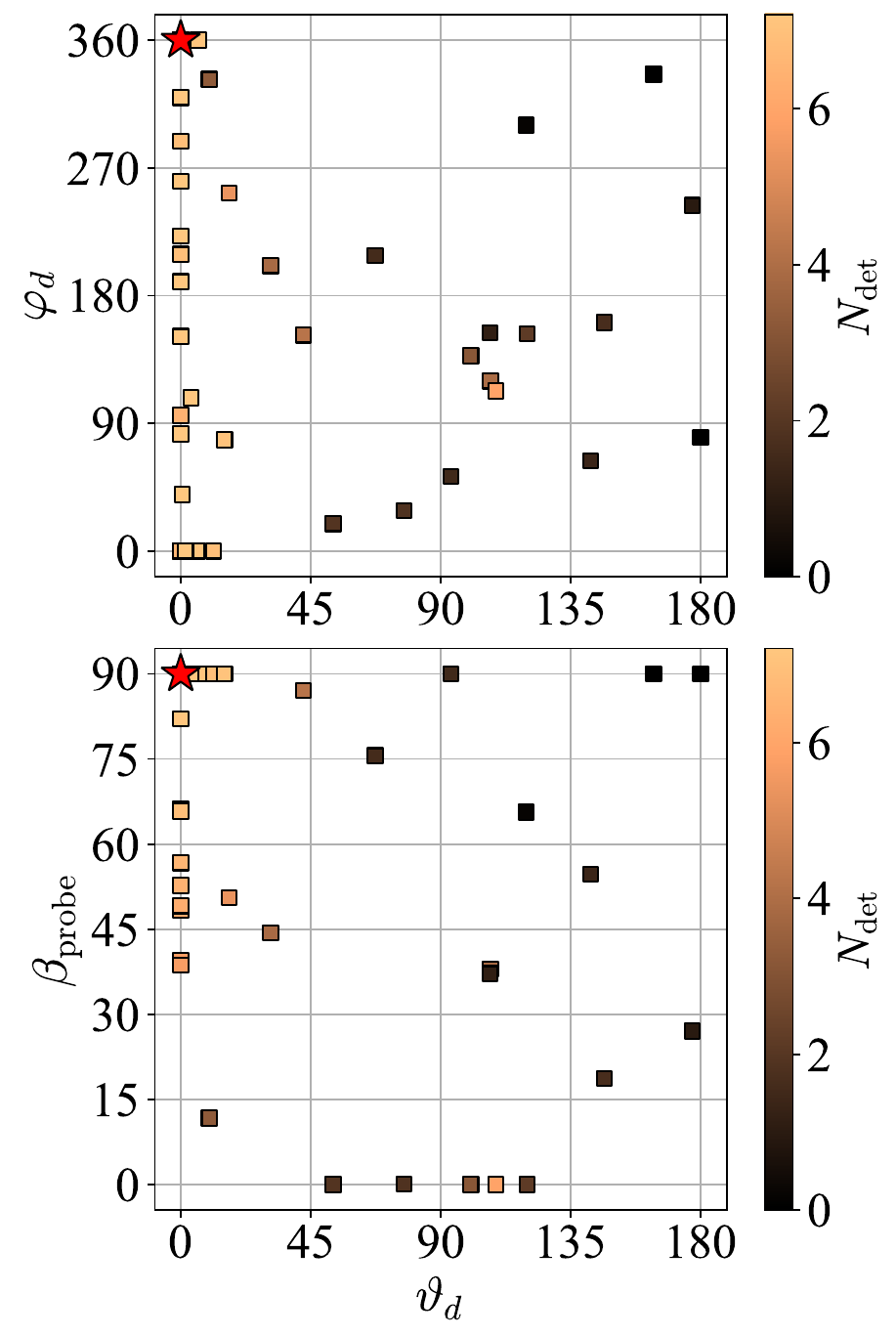}
    \caption{Optimization results in 3-dimensional parameter space:  $b$-dipole moment orientation ($\vartheta_d, \varphi_d$) and probe polarization ($\beta_{\rm probe}$). The optimization objective was to maximize the signal in a $10^{\circ} \times 10^{\circ}$ detector around $(\vartheta, \varphi) = (90^{\circ}, 180^{\circ})$. Probe and dipole pulse parameters are similar to Fig. \ref{fig:gg-gd}. Found optimum is $(\vartheta_d^*, \varphi_d^*, \beta_{\rm probe}^*) \approx (0^{\circ}, 360^{\circ}, 90^{\circ})$ achieving $N_{\rm{det}} \approx 7.2$. The optimum corresponds to $\mathbf{d}\: ||\: \mathbf{e}_z$ and $\mathbf{E_{\rm probe}\: || \:\mathbf{e}_y}$.}
    \label{fig:opt-dipole}
\end{figure}

To make sure that we found the optimal configuration, we ran a Bayesian optimization for 3 parameters: $b$-dipole moment orientation ($\vartheta_d, \varphi_d$) and probe polarization ($\beta_{\rm probe}$). All other parameters were similar to the ones from Fig. \ref{fig:gg-gd}. 
The results are shown on Fig. \ref{fig:opt-dipole} which suggests that the optimum lies around $(\vartheta_d^*, \varphi_d^*, \beta_{\rm probe}^*) \approx (0^{\circ}, 360^{\circ}, 90^{\circ})$ (this corresponds to $\mathbf{d}\: ||\: \mathbf{e}_z$ and $\mathbf{E_{\rm probe}\: || \:\mathbf{e}_y}$) achieving $N_{\rm{det}} \approx 7.2$. Note that for the special case of $\mathbf{d} \: \| \: \mathbf{e}_z$ with $\vartheta_d^* = 0^{\circ}$ the polar angle $\varphi_d^*$ is not defined.
With $\beta_{\rm probe} = 90^{\circ}$ this setup is equivalent to the one from Fig. \ref{fig:gg-gd} (b).

Overall, colliding a Gaussian probe with a dipole pulse of the same oscillation frequency results in a significant back-reflected signal, which is ``contaminated'' by other channels, including dipole self-scattering. We determine the optimal alignment between the dipole moment and the probe polarization to maximize the back-reflected signal. Unfortunately, a large dipole background renders this signature indiscernible: in the region of interest, the signal-to-background ratio is extremely low. In the next sections, we explore avenues to address this background issue.

\subsection{Multiple colliding pulses in a belt configuration}
\label{sec:belt}

In the previous section, we found that the collision of a Gaussian probe with a dipole pulse of the same frequency produces a sizable back-reflected signal. However, the dipole setup faces two serious issues: 1) a large dipole background makes the reflected signature not discernible, and 2) producing dipole pulses in the laboratory is very challenging. For a potential experimental measurement, it would be crucial to find a setup that both keeps the relevant spectral modes from the dipole pulse, achieving a sizable back-reflected signal, and does not have any background in the region of interest.

It was proposed in \cite{bulanov2010multiple} to use multiple colliding pulses arranged in a specific geometry to achieve high focusing. The multiple pulse configuration approximates the dipole pulse: the more pulses are taken, the closer the electromagnetic fields (and $k$-spectrum) resemble those of a dipole wave. Therefore, choosing a configuration with no pulses propagating close to the back-reflected direction might solve the background problem, making the back-reflected signature discernible.

\begin{figure}
    \centering
    \includegraphics[width=0.65\linewidth]{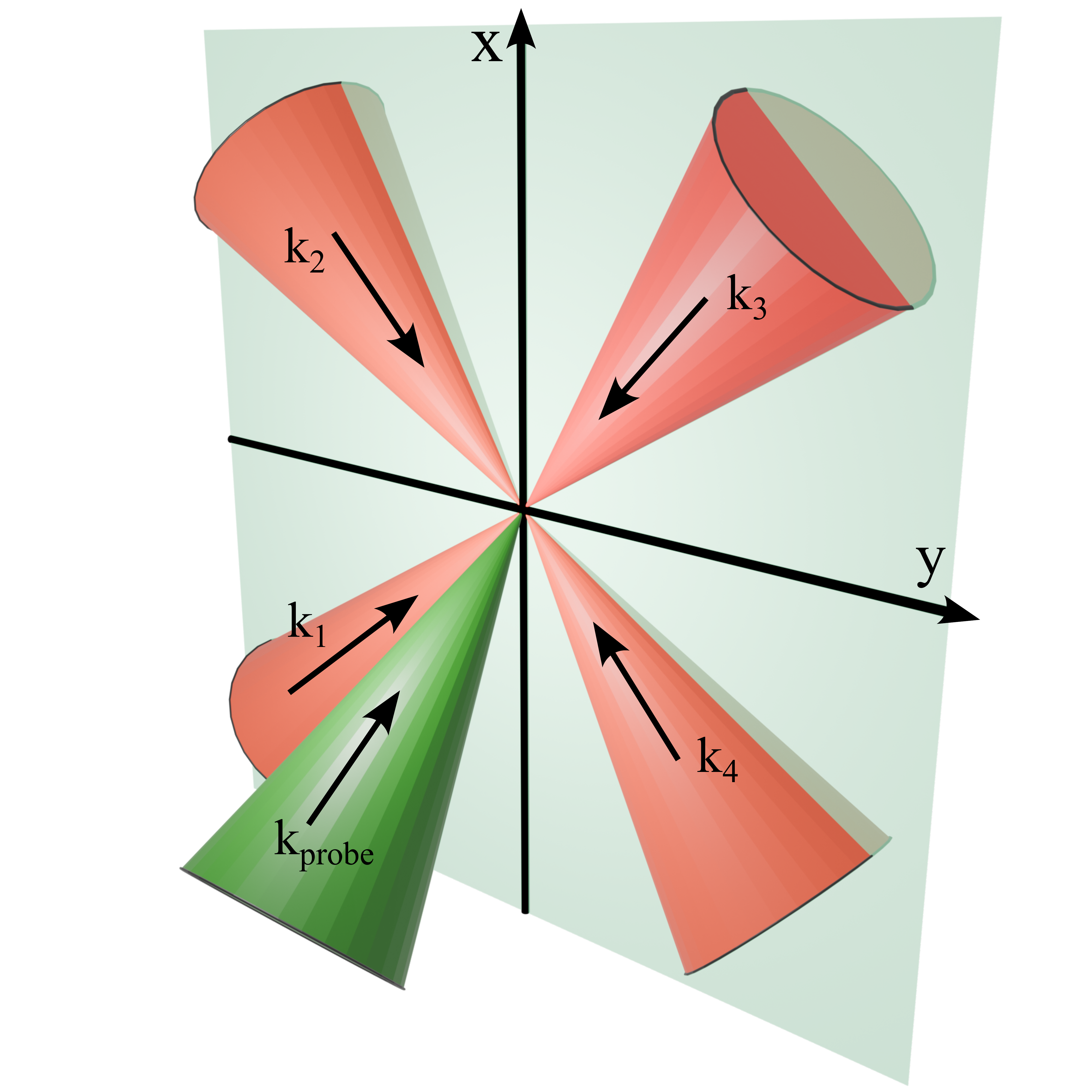}
    \caption{Schematic diagram of a belt collision scenario.}
    \label{fig:4-belt diagram}
\end{figure}

In this section, we consider a particular case of a multiple pulse focusing setup: four focused Gaussian pulses arranged in a belt configuration. Fig. \ref{fig:4-belt diagram} shows the geometry of the considered scenario: four pulses form a belt in the xy plane (the optical axis of each pulse is perpendicular to those of its neighbors), and a probe pulse hits the belt plane at $45^{\circ}$ in between the pulses described by wave vectors $\mathbf{k}_1$ and $\mathbf{k}_4$. From now on, we refer to the four pulses in a belt as ``belt configuration'' and the whole five pulse setup (with probe) as ``belt collision scenario''.

\begin{figure}
    \centering
    \includegraphics[width=0.95\linewidth]{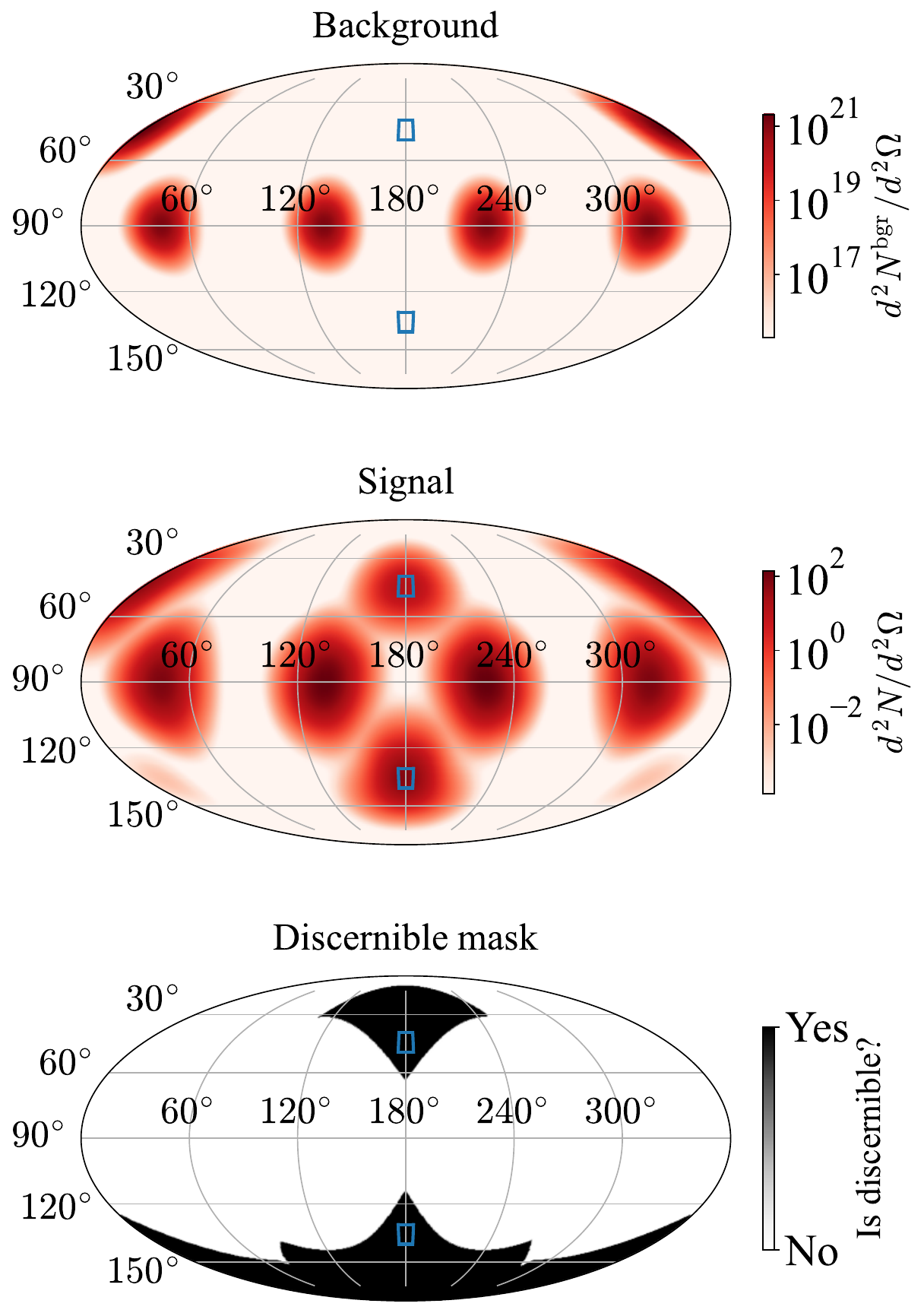}
    \caption{Angular background and signal photon spectrum and discernible mask from the collision of a Gaussian probe with a belt configuration. Probe parameters: $W = 20 \:\rm{J}$, $\lambda = 800\, \rm{nm}$, $w_0 = 2 \lambda$, $\tau_{\rm{FWHM}} = 20\, \rm{fs}$, $(\vartheta_{\rm probe}, \varphi_{\rm probe}) = (45^{\circ}, 0^{\circ})$, $\beta_{\rm probe}=0^{\circ}$. The belt consists of 4 pulses ($i=1..4$) with $W = 10\, \rm{J}$ each, $(\vartheta_{b,i}, \varphi_{b,i}) = (90^{\circ}, 90^{\circ} \times i - 45^{\circ})$ and polarization angles $\beta_1 = \beta_4 = 0^{\circ}, \beta_2 = \beta_3 = 90^{\circ}$ (other parameters are similar to the probe). Blue squares show $10^{\circ} \times 10^{\circ}$ detector regions. Signal around $(\vartheta, \varphi) = (135^{\circ}, 180^{\circ})$ is back-reflected with $N_{\rm{det}} \approx 1.2$, signal around $(\vartheta, \varphi) = (45^{\circ}, 180^{\circ})$ is forward-angle-reflected with $N_{\rm{det}} \approx 0.26$.}
    \label{fig:4-belt}
\end{figure}

Fig. \ref{fig:4-belt} shows the angular background and signal photon spectrum for the belt collision scenario. 
On the background plot, four distinct spots are clearly visible on the equator -- they correspond to the belt pulses; the remaining bright spot around ($\vartheta=45^{\circ}, \varphi=0^{\circ}$) is the probe pulse.
On the signal plot, in addition to the peaks related to the direction of the background pulses, we see three potentially discernible signatures: back-reflected ($\vartheta=135^{\circ}, \varphi=180^{\circ}$), back-angle-reflected ($\vartheta=135^{\circ}, \varphi=0^{\circ}$), and forward-angle-reflected ($\vartheta=45^{\circ}, \varphi=180^{\circ}$). For the chosen parameters in $10^{\circ} \times 10^{\circ}$ detectors we get $N_{\rm{det}}^{\rm{back-reflected}} \approx 1.2$ (signal-to-background ratio $\approx 1 \times 10^7$), $N_{\rm{det}}^{\rm{forward-angle-reflected}} \approx 0.26$ (signal-to-background ratio $\approx 3 \times 10^6$) and a negligible amount of back-angle-reflected signal.

\begin{figure}
    \centering
    \includegraphics[width=0.8\linewidth]{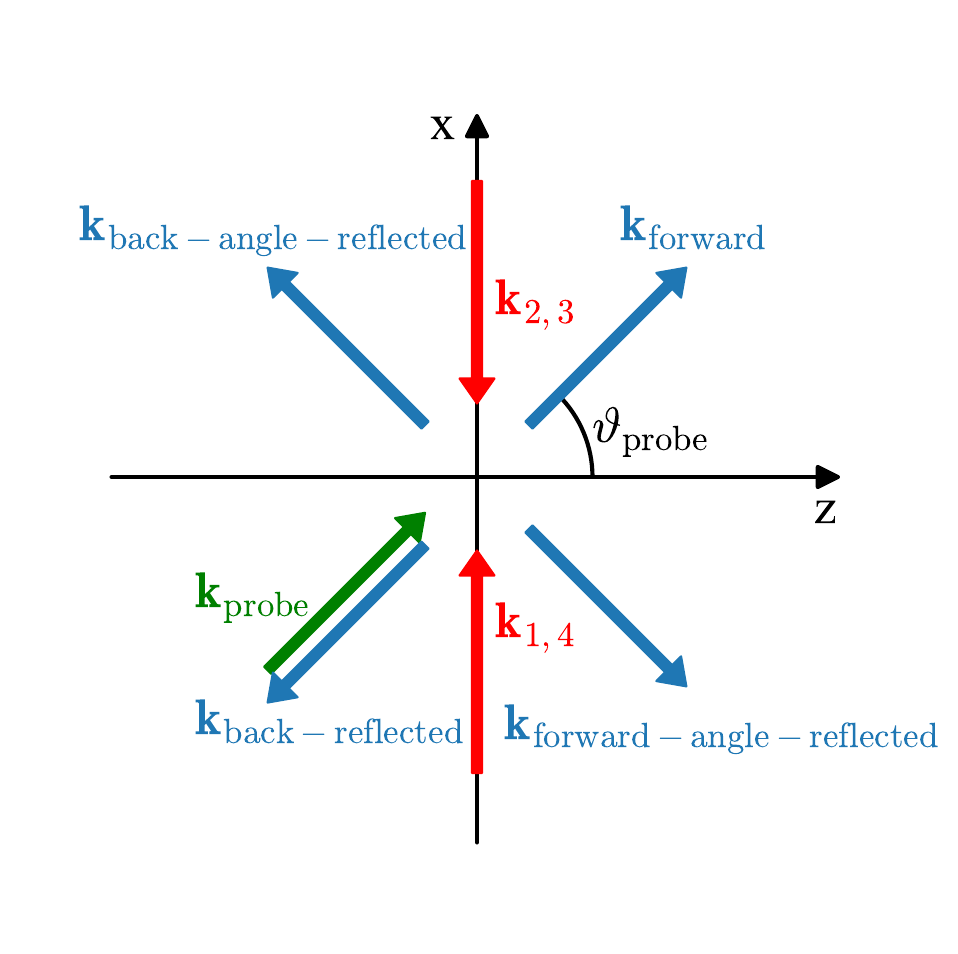}
    \caption{Schematic diagram of expected signals allowed by plane-wave energy-momentum conservation in a belt collision scenario.}
    \label{fig:belt-signal}
\end{figure}

To understand the origin of these discernible signatures, recall from Eq. \eqref{eq:channel-notation} that every signal channel scales cubically with the field. In multi-pulse collisions, any pair or triplet of participating pulses corresponds to a separate channel (self-scattering channels are suppressed for focused Gaussian pulses). For our setup in Fig. \ref{fig:4-belt diagram}, some viable combinations satisfying energy-momentum conservation are (p for probe): $S^{\mathrm{p}13}, S^{113}, S^{123}$, and so on. Each of the channels might result in a separate angularly or spectrally resolved signature. Since we are interested in an elastic back-reflected signature, we consider all pulses to have the same wavelength, so only angular signatures will be prominent. Moreover, this signature should be linear in the probe field, which limits the number of channels that might contribute to $\sum_{i,j=1}^4 S^{\mathrm{p} i j}$. From the plane wave estimates for our setup, we expect the following angular signatures (see Fig. \ref{fig:belt-signal}): 1) back-reflected ($\vartheta=135^{\circ}, \varphi=180^{\circ}$), 2) back-angle-reflected ($\vartheta=135^{\circ}, \varphi=0^{\circ}$), and 3) forward-angle-reflected ($\vartheta=45^{\circ}, \varphi=180^{\circ}$). The back-reflected signal is produced by ($+$ sign stands for the absorption of a photon, $-$ sign for its emission) $\mathbf{k}_{\rm signal} = \mathbf{k}_1 + \mathbf{k}_3 - \mathbf{k}_{\rm probe}$ and $\mathbf{k}_{\rm signal} = \mathbf{k}_2 + \mathbf{k}_4 - \mathbf{k}_{\rm probe}$, back-angle-reflected by $\mathbf{k}_{\rm signal} = \mathbf{k}_1 + \mathbf{k}_4 - \mathbf{k}_{\rm probe}$ and forward-angle-reflected by $\mathbf{k}_{\rm signal} = \mathbf{k}_1 + \mathbf{k}_2 - \mathbf{k}_{\rm probe}$ and $\mathbf{k}_{\rm signal} = \mathbf{k}_3 + \mathbf{k}_4 - \mathbf{k}_{\rm probe}$. The finite transverse and temporal extent of the pulses relaxes the plane-wave energy-momentum conservation condition. Observed signatures are consistent with this 4-wave mixing interpretation.

When the probe propagates closer to the xy plane ($\vartheta_{\rm probe} \rightarrow 90^{\circ}$), its interaction volume with the belt configuration increases, leading to a larger total back-reflected signal. At the same time, starting from some optimal $\vartheta_{\rm probe}^*$, the total discernible signal decreases as $\vartheta_{\rm probe} \rightarrow 90^{\circ}$ due to the background from belt pulses (see discernible mask on Fig. \ref{fig:4-belt}). We chose $\vartheta_{\rm probe} = 45^{\circ}$ to have a sizable signal and good signal-to-background ratio.
As for the azimuthal angle $\varphi_{\rm probe}$, we chose it so that the probe lies symmetrically between the pulses characterized by wave vectors $\mathbf{k}_1$ and $\mathbf{k}_4$. In this way, the signals coming from the fields $(\mathbf{k}_{\rm probe}, \mathbf{k}_1, \mathbf{k}_3)$ and $(\mathbf{k}_{\rm probe}, \mathbf{k}_2, \mathbf{k}_4)$ equally contribute to the same back-reflected signature (recall the channel discussion after Fig. \ref{fig:4-belt diagram}). Breaking this symmetry would result in different interaction volumes between the probe and pairs of pulses $(\mathbf{k}_1, \mathbf{k}_3)$ and $(\mathbf{k}_2, \mathbf{k}_4)$.

\begin{figure}
    \centering
    \includegraphics[width=0.8\linewidth]{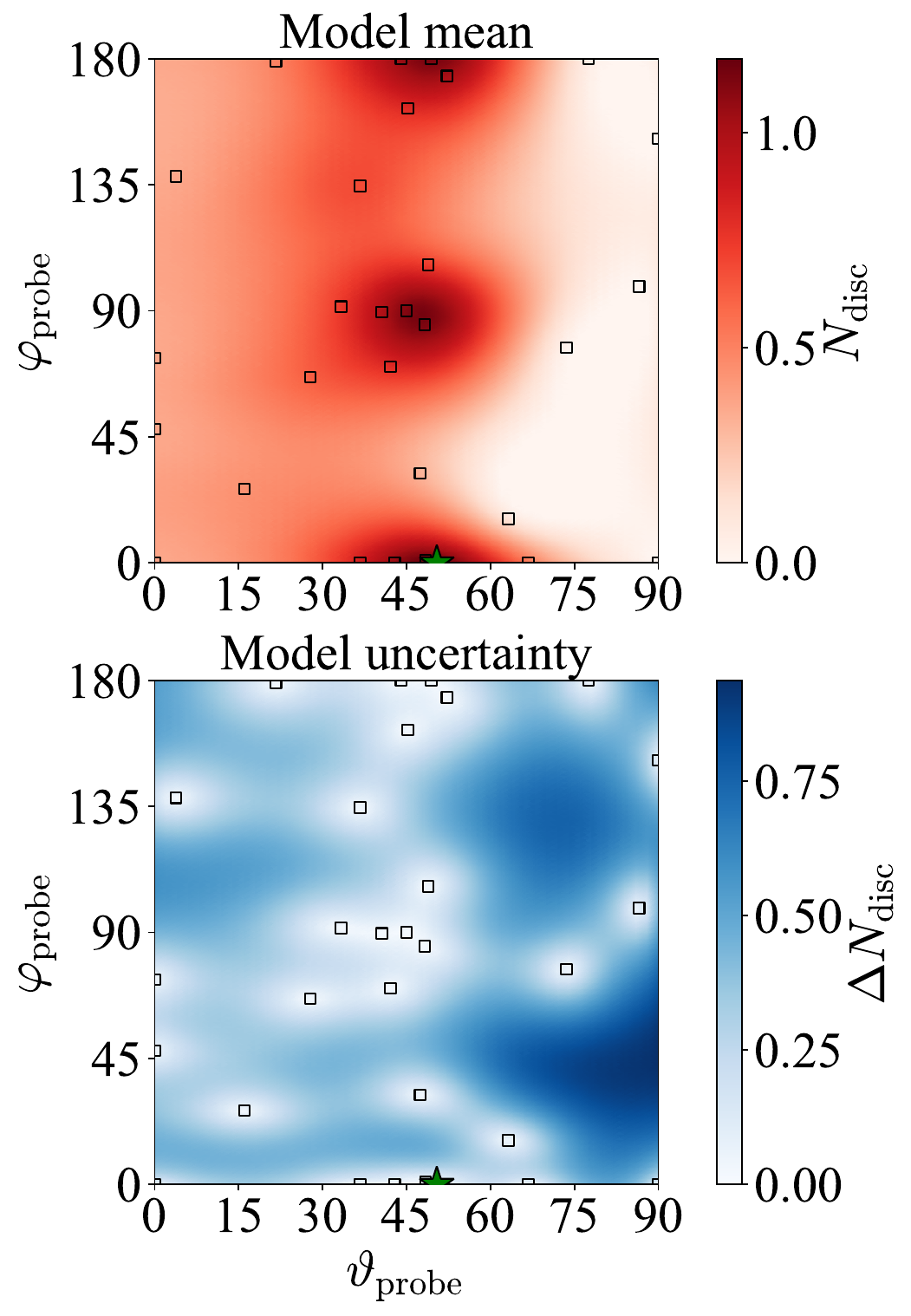}
    \caption{Optimization results in 2-dimensional parameter space: the propagation direction of the probe $(\vartheta_{\rm probe}, \varphi_{\rm probe})$. Colormaps correspond to (top) the mean and (bottom) 95\% confidence interval of the fitted Gaussian Process surrogate model. The optimization objective was to maximize the total discernible signal. The probe collides with a belt configuration consisting of four focused Gaussian pulses with polarizations $\beta_1 = \beta_2 = \beta_3 = \beta_4 = 0^{\circ}$. Other parameters are similar to Fig. \ref{fig:4-belt}. Found optimum is $(\vartheta_{\rm probe}^*, \varphi_{\rm probe}^*) \approx (50.4^{\circ}, 0^{\circ})$ achieving $ N_{\rm{disc}} \approx 1.2$. The top plot shows us that there are three equivalent configurations with $\vartheta_{\rm probe} \approx 50.4^{\circ}$: 1) $\varphi_{\rm probe} = 0^{\circ}$ $\leftrightarrow$ probe propagates between pair of pulses described by $(\mathbf{k}_4,\mathbf{k}_1)$, 2) $\varphi_{\rm probe} = 90^{\circ}$ $\leftrightarrow$ between $(\mathbf{k}_1,\mathbf{k}_2)$, 3) $\varphi_{\rm probe} = 180^{\circ}$ $\leftrightarrow$ between $(\mathbf{k}_2,\mathbf{k}_3)$. These configurations result in the same total discernible signal, and optimization converged to one of them. When $\vartheta_{\rm probe}$ is too large, no sizable discernible signal could be detected. The bottom plot shows that around the observation points the uncertainty is zero (as expected), while in unexplored parameter regions it is much larger.}
    \label{fig:opt-4-belt-probe}
\end{figure}

To test that we chose a reasonable probe orientation, we performed a 2d Bayesian optimization for the probe's propagation direction $(\vartheta_{\rm probe}, \varphi_{\rm probe})$. 
The results are shown in Fig. \ref{fig:opt-4-belt-probe}. Note that this optimization plot differs from others: here, the color corresponds to the mean and variance of the Gaussian Process surrogate model fitted to the observations. When the color does not overlap with observations (squares), it does not represent the ``true'' amount of the discernible signal obtained from simulations but rather the model's ``belief'' (mean and 95\% confidence interval) of what the value would be. For a 2-dimensional parameter space, it could be represented nicely, but for higher-dimensional parameter spaces, it becomes challenging for visualization and interpretation. 
The results suggest that the maximal discernible signal is achieved for the configuration $(\vartheta_{\rm probe}^*, \varphi_{\rm probe}^*) \approx (50.4^{\circ}, 0^{\circ})$, which is close to our setup. Note that the symmetric configurations with $(\vartheta_{\rm probe}^*, \varphi_{\rm probe}^*) \approx (50.4^{\circ}, 90^{\circ})$ and $(50.4^{\circ}, 180^{\circ})$ result in the same amount of discernible signal. For $\vartheta_{\rm probe} \rightarrow 90^{\circ}$ the discernible signal almost vanishes. Discernibility in this case (and optimal $\vartheta_{\rm probe}^*$) is closely linked to the focusing of background pulses: the setups with weaker focusing would have larger discernible area (recall discernible map from Fig. \ref{fig:4-belt}) producing sizable discernible signal even for larger $\vartheta_{\rm probe}$ and shifting the optimal collision angle closer to the xy plane ($\vartheta_{\rm probe}^* \rightarrow 90^{\circ}$ as $w_0 \rightarrow \infty$). On the uncertainty plot, the areas around observations have low variance, which increases the further one goes in the ``unexplored'' area.

\begin{figure}
    \centering
    \includegraphics[width=0.85\linewidth]{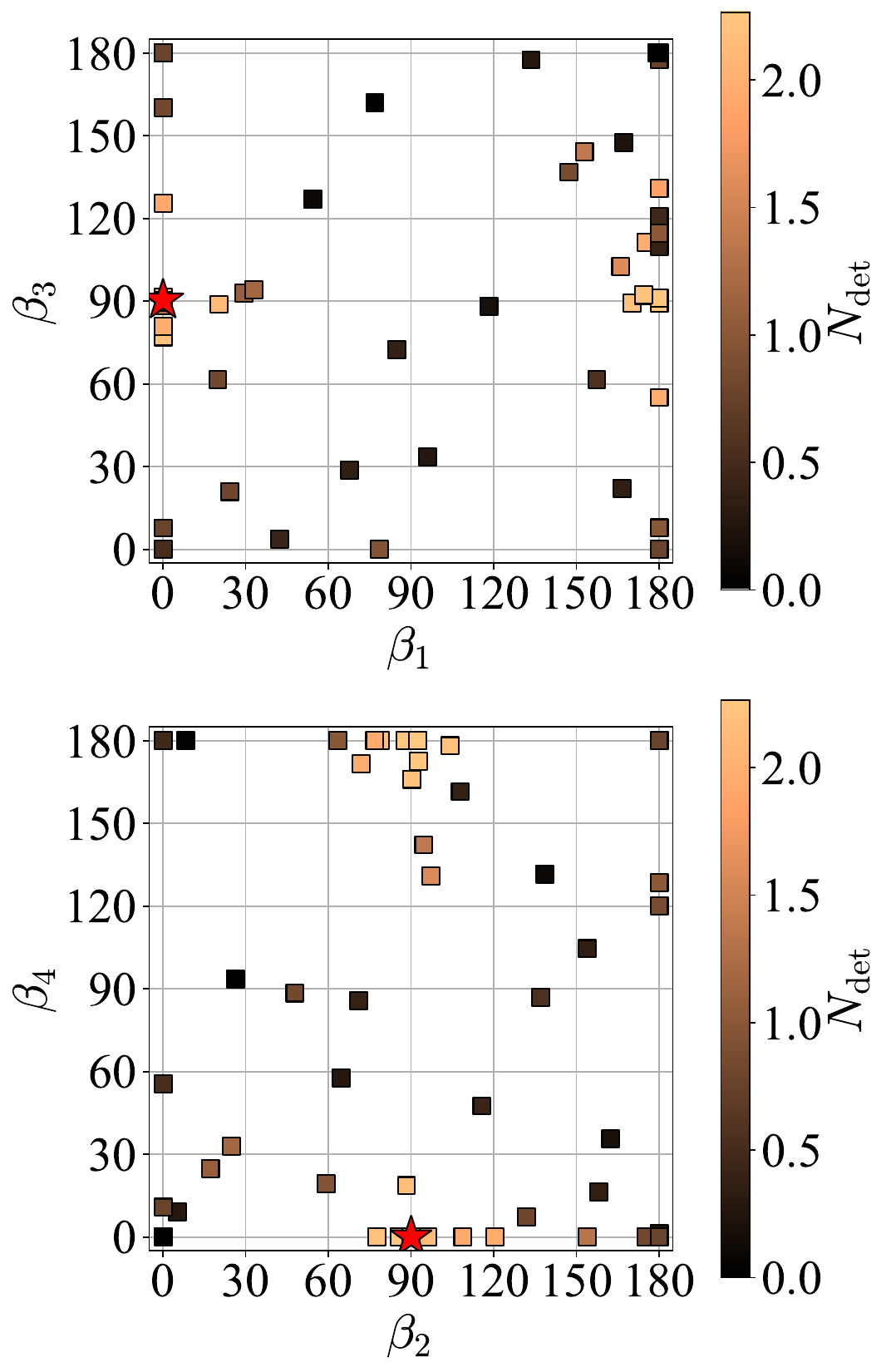}
    \caption{Optimization results in 4-dimensional parameter space: the linear polarization angle of the different belt pulses $(\beta_1, \beta_2, \beta_3, \beta_4)$. The optimization objective was to maximize the number of photons in a $10^{\circ} \times 10^{\circ}$ detector around $(\vartheta, \varphi) = (90^{\circ}, 180^{\circ})$. The probe collides with a belt configuration consisting of 4 focused Gaussian pulses at an angle of $\vartheta_{\rm probe} = 90^{\circ}$, the polarization of the probe is fixed to $\beta_{\rm probe} = 0^{\circ}$. Other parameters are similar to Fig. \ref{fig:4-belt}. Found optimum is $(\beta_1^*, \beta_2^*, \beta_3^*, \beta_4^*) \approx (0^{\circ}, 90^{\circ}, 90^{\circ}, 0^{\circ})$ achieving $ N_{\rm{det}} \approx 2.3$.}
    \label{fig:opt-4-belt-polarization}
\end{figure}

To find the optimal polarization configuration between the belt pulses, we launched a 4d Bayesian optimization. The results are shown in Fig. \ref{fig:opt-4-belt-polarization}. They suggest that for a fixed polarization of the probe $\beta_{\rm probe}$, several equivalent optimal configurations result in the same amount of detected signal (e.g., $(\beta_1, \beta_2, \beta_3, \beta_4) = (0^{\circ}, 90^{\circ}, 90^{\circ}, 0^{\circ})$ and $(\beta_1, \beta_2, \beta_3, \beta_4) = (180^{\circ}, 90^{\circ}, 90^{\circ}, 180^{\circ})$). Changing probe polarization shifts the optimum (see Appendix \ref{sec:pol-config} for a more detailed discussion in the context of three pulse collisions). In this optimization setting, we chose the probe to lie in the belt plane (which results in a larger total signal compared to our setup on Fig. \ref{fig:4-belt}) to minimize the computational cost.

\begin{figure}
    \centering
    \includegraphics[width=0.85\linewidth]{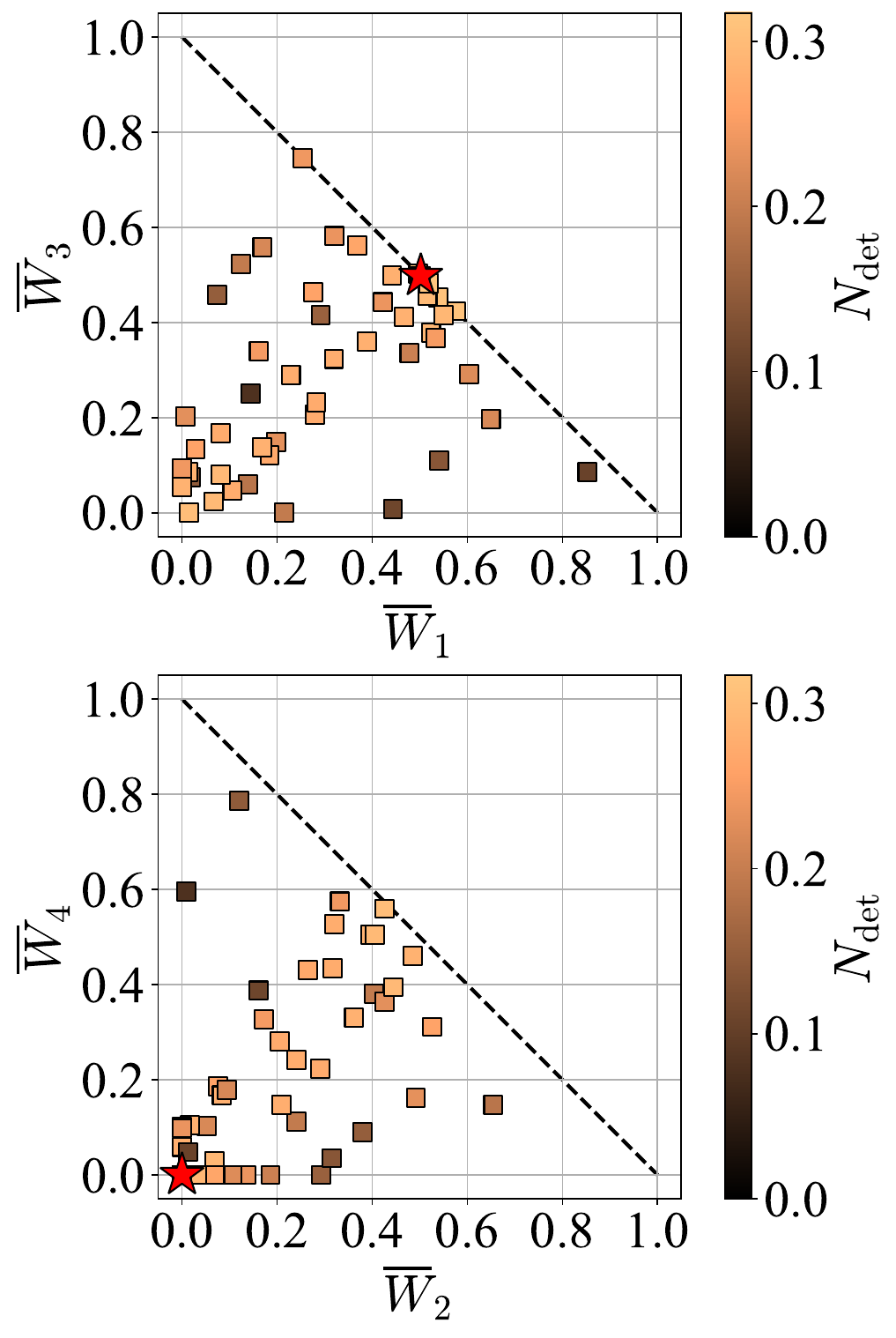}
    \caption{Optimization results in 3-dimensional parameter space: energy fraction of three pulses in the belt configuration ($\overline{W}_1, \overline{W}_2, \overline{W}_3$). The energy fraction of the fourth pulse is $\overline{W}_4 = 1 - \overline{W}_1 - \overline{W}_2 - \overline{W}_3$. The total energy budget for the belt pulses was fixed to $W_{\rm{pump}} = 40\, \rm{J}$. The optimization objective was to maximize the signal in a $10^{\circ} \times 10^{\circ}$ detector around $(\vartheta, \varphi) = (135^{\circ}, 180^{\circ})$. The polarization of pulses is $\beta_1 = \beta_2 = \beta_3 = \beta_4 = 0^{\circ}$, other parameters are similar to Fig. \ref{fig:4-belt}. Found optimum is $(\overline{W}_1^*, \overline{W}_2^*, \overline{W}_3^*, \overline{W}_4^*) \approx (0.5, 0, 0.5, 0)$ achieving $N_{\rm{det}} \approx 0.32$.}
    \label{fig:opt-4-belt-energy}
\end{figure}

Recall from Eq. \eqref{eq:channel-notation} that every signal channel scales cubically with the field, and the back-reflected channel is linear in the probe field and quadratic in the pump field. For a fixed energy budget $W_0$, the maximal total back-reflected signal is achieved when the probe gets $W_{\rm probe} = 1/3 \: W_0$ and the pump gets $W_{\rm{pump}} = 2/3 \: W_0$.
For our setup, we choose these optimal values where $W_{\rm{pump}}$ is equally distributed between four pulses, resulting in $W_{\rm{belt\: pulse}} = 1/6 \: W_0$.
To test that such energy distribution indeed maximizes the back-reflected signal, we launched another Bayesian optimization where we had a fixed energy budget for belt pulses equal to $W_{\rm{pump}}$, and the energies of three pulses were optimized parameters ($W_1, W_2, W_3$) with a constraint $W_1 + W_2 + W_3 \leq W_{\rm{pump}}$ and $W_4 = W_{\rm{pump}} - W_1 - W_2 - W_3$. Fig. \ref{fig:opt-4-belt-energy} shows the optimization results, which suggest that removing two pulses from the belt configuration (redistributing $W_{\rm{pump}}$ between only two remaining ones) leads to an even larger back-reflected signal.
The optimization converged to the configuration where $W_2$ and $W_4$ are removed, but we see equally bright trials when $W_1$ and $W_3$ are removed. Both of these three pulse configurations are equivalent.

Overall, replacing a dipole pulse with a belt configuration produces a sizable \textit{discernible} back-reflected signal, improving upon the previous setup. We provide a channel analysis of signals from the belt configuration and the optimal values for the probe orientation, polarization of the belt pulses, and energy distribution between them. In particular, energy distribution optimization shows that removing two pulses from the five-pulse belt setup leads to an even larger back-reflected signal.
From an experimental point of view, the more pulses collide, the harder it is to precisely overlap them temporally and spatially (due to jitter), making a three-pulse setup a favorable choice. In the next section, we will explore this setup.

\subsection{Three pulse collision}
\label{sec:3-beams}

In the previous section, we found that the three pulse setup yields a larger amount of total back-reflected signal in the region of interest than the five pulse belt setup.
Three pulse setups were extensively studied in the light-by-light scattering context: first analytic estimates \cite{varfolomeev1966induced, rozanov1993four}, setups with different optical frequencies \cite{Bernard:2000ovj, Lundstrom:2005za, lundin2006analysis, Gies:2017ezf, King:2018wtn, zhang2025computational, Rinderknecht:2025zgu}, setups with an x-ray pulse \cite{DiPiazza:2006pr, King:2018wtn, Ahmadiniaz:2022nrv}, setups with an OAM pulse \cite{Aboushelbaya:2019ncg}, photon merging and splitting \cite{Gies:2016czm}, general analytic study \cite{Berezin:2024fxt}. The motivation of these setups is to leverage the third pulse to produce the signal which would differ from the background by its angular, frequency, polarization, or OAM properties, or a combination of them.
In this section, we focus on a subset of three pulse setups: 1) every pulse has the same frequency in the optical range, and 2) a specific planar geometry resulting from our belt setup. 

\begin{figure}
    \centering
    \includegraphics[width=0.65\linewidth]{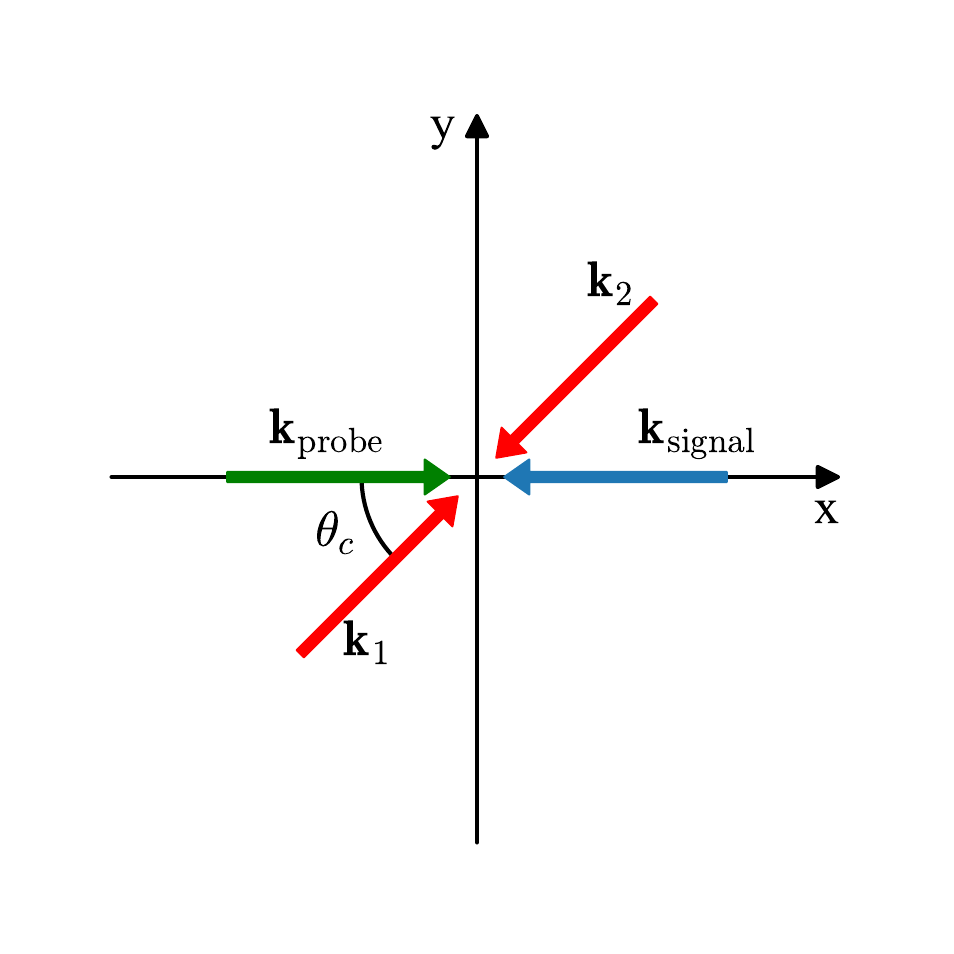}
    \caption{Schematic diagram of a three pulse collision scenario.}
    \label{fig:diagram_3_beams}
\end{figure}

Our planar geometry for three pulses collision is shown in Fig. \ref{fig:diagram_3_beams}. Two counter-propagating pulses form the pump, and a probe pulse propagates at $\theta_c$ angle to it. The back-reflected signal is formed by $\mathbf{k}_{\rm{signal}} = - \mathbf{k}_{\rm probe} + \mathbf{k}_1 + \mathbf{k}_2$. Note that this configuration results from Fig. \ref{fig:4-belt diagram} by removing two counter-propagating belt pulses (e.g. $\mathbf{k}_2, \mathbf{k}_4$), relabeling the remaining belt pulses to $\mathbf{k}_1, \mathbf{k}_2$ (e.g. $\mathbf{k}_1 \rightarrow \mathbf{k}_1, \mathbf{k}_3 \rightarrow \mathbf{k}_2$) and rotating the coordinate frame so all three pulses lie on the equator (this is always possible for our geometry since two pulses are counter-propagating). 

The collision setting we consider closely resembles those in \cite{varfolomeev1966induced, rozanov1993four, Gies:2016czm, Rinderknecht:2025zgu}. The difference is that in our simulations, the background fields are exact solutions of linear Maxwell equations (without simplifying assumptions regarding the electromagnetic field profile). Additionally, we focus purely on the quasi-elastic back-reflected signature (signal is at the probe's frequency and in the direction opposite to the probe propagation). It should be noted that it might be experimentally challenging to realize the setup with two exactly counter-propagating pulses (as considered here). For pump pulses propagating at some angle, the energy-momentum conservation would be modified, requiring the probe to have a different frequency in order to produce a sizable signal in the back-reflected direction. Eventually, this would correspond to the setups discussed in \cite{Bernard:2000ovj,lundin2006analysis,zhang2025computational}. 

\begin{figure}
    \centering
    \includegraphics[width=0.95\linewidth]{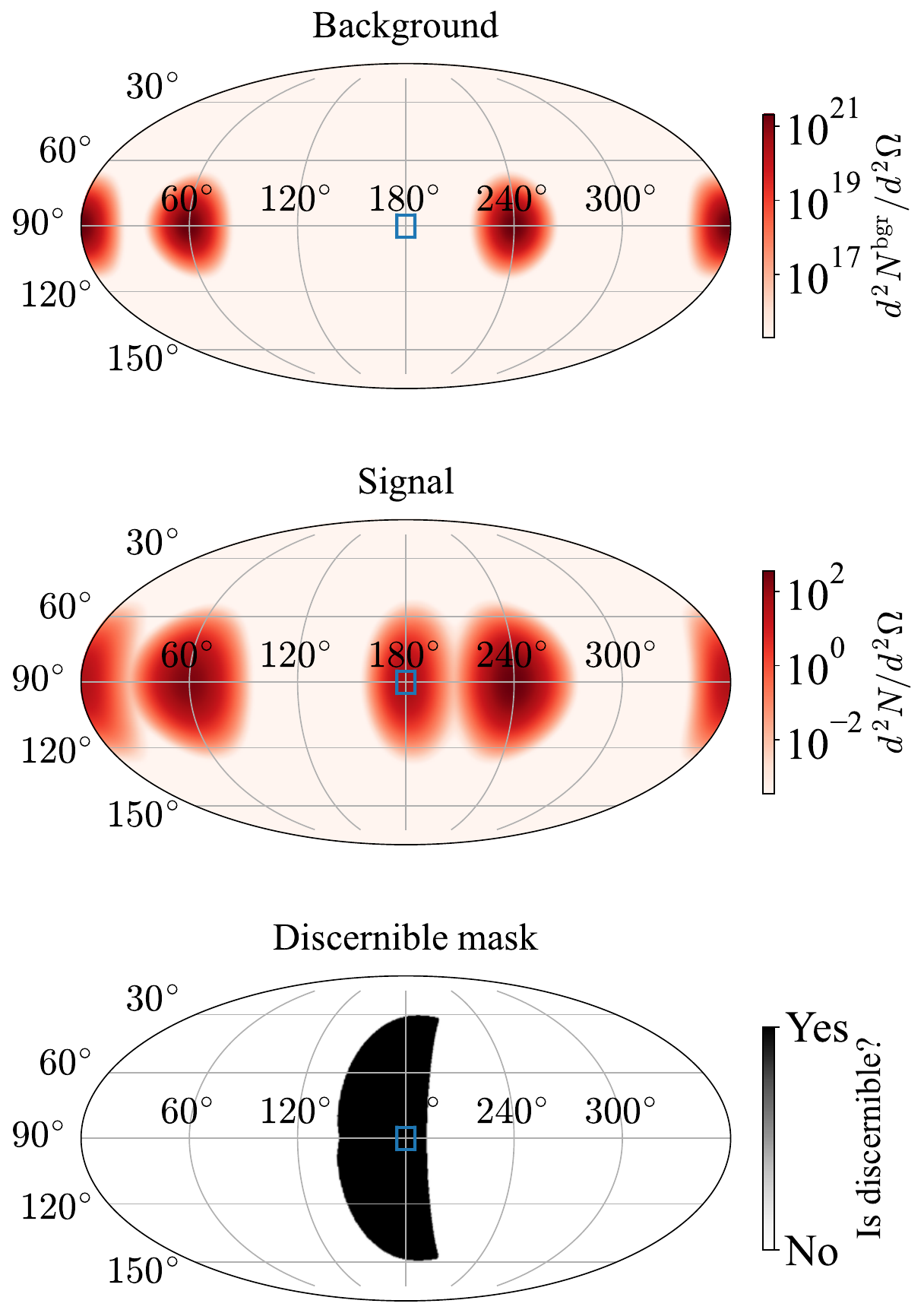}
    \caption{Angular background and signal photon spectrum and discernible mask from the collision of three Gaussian pulses. All pulses are characterized by $W = 20\, \rm{J}$, $\vartheta = 90^{\circ}$, $\lambda = 800\, \rm{nm}$, $\tau_{\rm{FWHM}} = 20\, \rm{fs}$, $w_0 = 2\lambda$ and $(\varphi_{\rm probe}, \varphi_1, \varphi_2) = (0^{\circ}, 60^{\circ}, 240^{\circ}), (\beta_{\rm probe}, \beta_1, \beta_2) = (0^{\circ}, 0^{\circ}, 90^{\circ})$. Blue square shows $10^{\circ} \times 10^{\circ}$ detector region with $N_{\rm{det}} \approx 1.5.$}
    \label{fig:3-beams}
\end{figure}

Fig. \ref{fig:3-beams} shows the angular background and signal photon spectrum for this scenario.
The angles $(\varphi_{\rm probe}, \varphi_1, \varphi_2) = (0^{\circ}, 60^{\circ}, 240^{\circ})$ were chosen to agree with our initial belt setup.
On the background plot, three spots correspond to three pulses, and on the signal plot, the extra spot at $(\vartheta, \varphi) = (90^{\circ}, 180^{\circ})$ corresponds to the back-reflected signature. The chosen configuration results in $N_{\rm{det}} \approx 1.5$ signal photons in the detector region (signal-to-background ratio $\approx 3 \times 10^7$). 
Compared to the belt setup, the absence of two additional background pulses increases the potential discernible area. As a result, by moving the optical axis of pump pulses closer to the optical axis of the probe it is possible to slightly increase the signal at the expense of signal-to-background ratio (e.g., the setup with $(\varphi_{\rm probe}, \varphi_1, \varphi_2) = (0^{\circ}, 55^{\circ}, 235^{\circ})$ results in $N_{\rm{det}} \approx 1.8$ and signal-to-background ratio $\approx 3 \times 10^3$).

We chose the optimal polarization angles from the previous setup. To test if this is reasonable, we also provide a simple plane-wave-based analytic estimate for the polarization prefactor in the vacuum emission amplitude (see Appendix \ref{sec:pol-config}). The analytic results suggest that the optimal polarization configuration doesn't depend on $\theta_c$ (a smaller $\theta_c$ increases the prefactor but does not affect the location of the optimum $(\beta_{\rm probe}^*, \beta_1^*, \beta_2^*)$)
and for each fixed $\beta_{\rm probe}$ the optimal $\beta_1^* = \beta_{\rm probe}$, and $\beta_1^* + \beta_2^* = 90^{\circ} + 180^{\circ} n$ with $n = 0$ for $\beta_{\rm probe} < 90^{\circ}$ and $n = 1$ for $ 90^{\circ} < \beta_{\rm probe} < 180^{\circ}$.

Overall, we studied a subset of three pulse setups, which is motivated by our belt setup and produces a sizable discernible back-reflected signal. Being more experimentally accessible than other considered scenarios (dipole and belt), it still has a few challenges. First of all, since our signals are at the frequency of driving lasers, it would be especially crucial not to allow free electrons in the region between the interaction point and the field-of-view of the detector \cite{Rinderknecht:2025zgu}. Secondly, timing and pointing jitter between three pulses might result in a shot-to-shot signal variability \cite{Rinderknecht:2025zgu}.

\section{Conclusions}
\label{sec:conclusions}

Dipole pulses are a particularly interesting field configuration with a high focusing efficiency that has been argued to have benefits for $e^+e^-$ pair creation \cite{Gonoskov:2013aoa, Gonoskov:2016ynx}. Motivated by this, a dipole pulse was considered to be a promising candidate for the real-world implementation of quantum reflection that could yield a sizable signal. First numerical estimates for light-by-light scattering with dipole pulses were provided, focusing on a signal observed in the back-reflected direction relative to the probe. The optimal orientation of dipole moment and probe polarization to maximize the back-reflected signal was determined in an equal-frequency setting, and it was found that the signal mainly originates from a specific 4-wave mixing channel rather than from pure quantum reflection from localized inhomogeneous potentials. Unfortunately, despite the dipole configuration having a solid amount of reflected signal, it remains indistinguishable from the dominant dipole background. 

To address the background issue, we study the belt setup (with four pulses forming the belt), which is a particular case of a finite number of pulses approximation of the dipole pulse. We show that such a configuration does result in a discernible back-reflected signature. The optimal probe propagation direction maximizing the discernible signal and the polarization configuration enhancing the back-reflected signal were determined. Following our optimization results for the energy distribution between the belt pulses, we study the planar three pulse geometry, which yields an even larger back-reflected signal than the belt setup. Among all considered scenarios, the three pulse setup appears to be the most experimentally accessible, though it remains challenging to implement in practice.

\section*{Data Availability}
We developed and used the package \textit{quvac} \cite{quvac} to obtain the results for this article (version 0.1.1 is available \cite{quvac-zenodo}). 
The data that support the findings of this article are openly available \cite{data-repository}.

\acknowledgements
The authors acknowledge the usage of the HPC cluster “DRACO” of the University of Jena for obtaining the numerical results presented in this article.

The research leading to the presented results received additional funding from the European Regional Development Fund and the State of Thuringia via Thüringer Aufbaubank TAB (Contract No.~2019 FGI 0013). This work has been funded also by the Deutsche Forschungsgemeinschaft (DFG) under Grants No. 416607684, No. 416702141, and No. 416708866 within
the Research Unit FOR2783/2.

\bibliography{references}

\clearpage
\onecolumngrid
\appendix

\section{Channels in the dipole setup}
\label{sec:dipole-channels}

\begin{figure}[h!]
    \centering
    \includegraphics[width=1\linewidth]{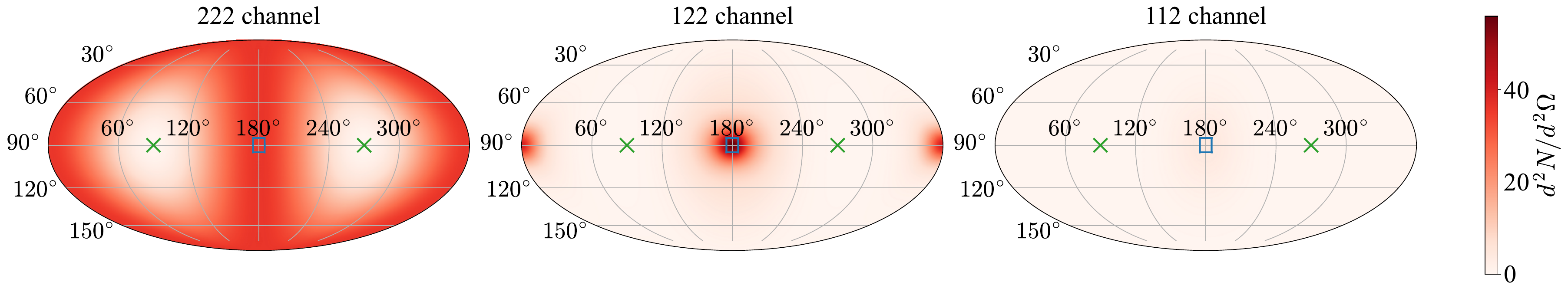}
    \caption{Angular signal photon spectrum for particular channels from the collision of a Gaussian probe with a dipole pulse. Setup parameters are similar to Fig. \ref{fig:gg-gd} (b). Blue square shows $10^{\circ} \times 10^{\circ}$ detector region. (Left) 222 channel cubic in the dipole field corresponds to dipole self-emission and yields $N_{\rm tot} \approx 289,\: N_{\rm det} \approx 1.15$. (Middle) 122 channel linear in the probe field corresponds to the back-reflected signature we look for and yields $N_{\rm tot} \approx 38,\: N_{\rm det} \approx 1.62$. (Right) 112 channel linear in the dipole field yields $N_{\rm tot} \approx 4.4,\: N_{\rm det} \approx 0.13$.}
    \label{appendix-fig:dipole-channels}
\end{figure}

Fig. \ref{appendix-fig:dipole-channels} shows the signal spectra originating from 222, 122 and 112 channels in the collision of a Gaussian probe (1) with a dipole pulse (2). 122 and 112 plots were calculated using the vacuum emission amplitude linearized in one of the fields (this functionality is available in \textit{quvac}). Note that 222 and 122 channels are dominant, and 122 channel has a larger peak in the back-reflected direction than in the forward one. In the total vacuum emission signal, the interference terms also play a significant role, especially the interference between 222 and 122 channels.

\section{Polarization configuration for three pulse setup}
\label{sec:pol-config}

Here, we outline the procedure we used to estimate the polarization prefactor in the emission amplitude.

Following \cite{Karbstein:2015qwa}, the single photon emission amplitude is given by
\begin{equation}
    S_{(p)}(\mathbf{k}) \approx \frac{1}{45} \frac{m^2}{8 \pi^2} \frac{i e}{\sqrt{2k}} \left( \frac{e}{m^2} \right)^3 \int d^4 x\: e^{i k x} \left[ 4 \mathcal{F}(x)\: F_{\mu\nu}(x) + 7 \mathcal{G}(x) \: {}^{*} F_{\mu\nu}(x) \right] \: \hat{f}_{(p)}^{\mu\nu},
\end{equation}
where $\hat{f}_{(p)}^{\mu\nu} = k^{\mu} \epsilon^{*\nu}_{(p)}(k) - k^{\nu} \epsilon^{*\mu}_{(p)}(k)$ is the normalized field strength tensor of the emitted photon in the momentum space.

In spherical coordinates $\hat{\mathbf{k}} = (\cos\varphi \sin\vartheta, \sin\varphi \sin\vartheta, \cos\vartheta)$ and photon polarization modes are given by $\epsilon_{(p)}^{\mu} = (0, \mathbf{e}_p)$
\begin{align}
    \mathbf{e}_1 = \begin{pmatrix}
    \cos\varphi \cos\vartheta \\
    \sin\varphi \cos\vartheta \\
    -\sin\vartheta
    \end{pmatrix},\:
    \mathbf{e}_2 = \begin{pmatrix}
    -\sin\varphi \\
    \cos\varphi \\
    0
    \end{pmatrix},
\end{align}

We specify the background fields as plane-waves with defined k-vector and directions of electric and magnetic fields (for $\mathbf{k} = \mathbf{e}_z$, the polarization $\beta = 0^{\circ}$ corresponds to the $\hat{\mathbf{E}} = \mathbf{e}_x$ and $\hat{\mathbf{B}} = \mathbf{e}_y$). The geometry of the collision is similar to Fig. \ref{fig:diagram_3_beams}. After the estimation of the background field factor, we collect the terms that are linear in the amplitude of each field (terms containing $E_{1} E_{2} E_{\rm probe}$). Then we choose k-vector corresponding to background reflection with respect to the probe: $\varphi \rightarrow 180^{\circ}, \vartheta \rightarrow 90^{\circ}$. Leaving only the factor depending on the polarization angles, we get:

\begin{align}
\label{eq:pol-prefactor}
    P = &- 2838 \{\cos 2
   \left(\beta_1+\beta_{\rm probe}\right) + \cos 2 \left(\beta_2+\beta_{\rm probe}\right)\} + 4 \{ 484 \cos 2 \left(\beta_1-\beta_{\rm probe}\right)+973 \} \cos \left(2 \theta_c\right) + \nonumber \\
   &5808 \cos 2 \left(\beta_1-\beta_{\rm probe}\right) + 1936 \cos 2 \left(\beta_2-\beta_{\rm probe}\right) \{\cos 2 \theta_c + 3\} - 1056 \cos 2 \left(\beta_1+\beta_2\right) \{ \cos 2 \theta_c + 7 \} - \nonumber \\
   &66 \{ \cos 2 \left(\beta_1+\beta_{\rm probe}\right)+\cos 2
   \left(\beta_2+\beta_{\rm probe}\right)\} \{ 20 \cos 2 \theta_c + \cos 4 \theta_c \} + 968 \cos 2 \left(\beta_1-\beta_2\right) \sin ^4\theta_c - \\
   &176 \sin \left(\beta_1-\beta_2\right) \{ 88 \sin \left(\beta_1+\beta_2-2 \beta_{\rm probe}\right) \cos \theta_c + 3 \sin \left(\beta_1+\beta_2+2 \beta_{\rm probe}\right) [15 \cos \theta_c + \cos 3 \theta_c] \} + \nonumber \\
   &139 \cos 4 \theta_c + 13761 \nonumber
\end{align}

\begin{figure}[h]
    \centering
    \includegraphics[width=0.95\linewidth]{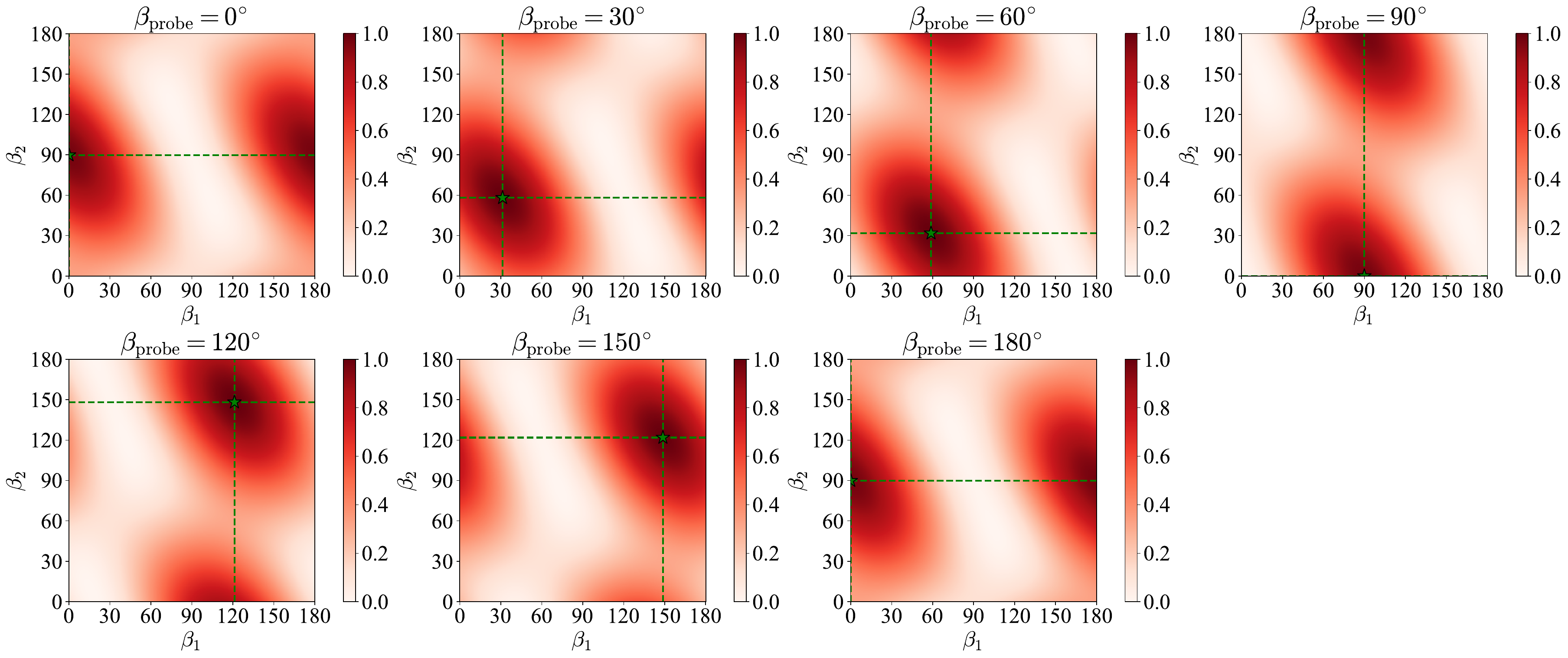}
    \caption{Polarization prefactor (from Eq. \ref{eq:pol-prefactor}) dependence on the polarization of individual beams ($\beta_1, \beta_2, \beta_{\rm probe}$) for a fixed collision angle $\theta_c = 45^{\circ}$. The maximal prefactor value was normalized to 1, with a normalizing factor of $\approx 38072$.}
    \label{appendix-fig:pol-45}
\end{figure}

\begin{figure}[h]
    \centering
    \includegraphics[width=0.95\linewidth]{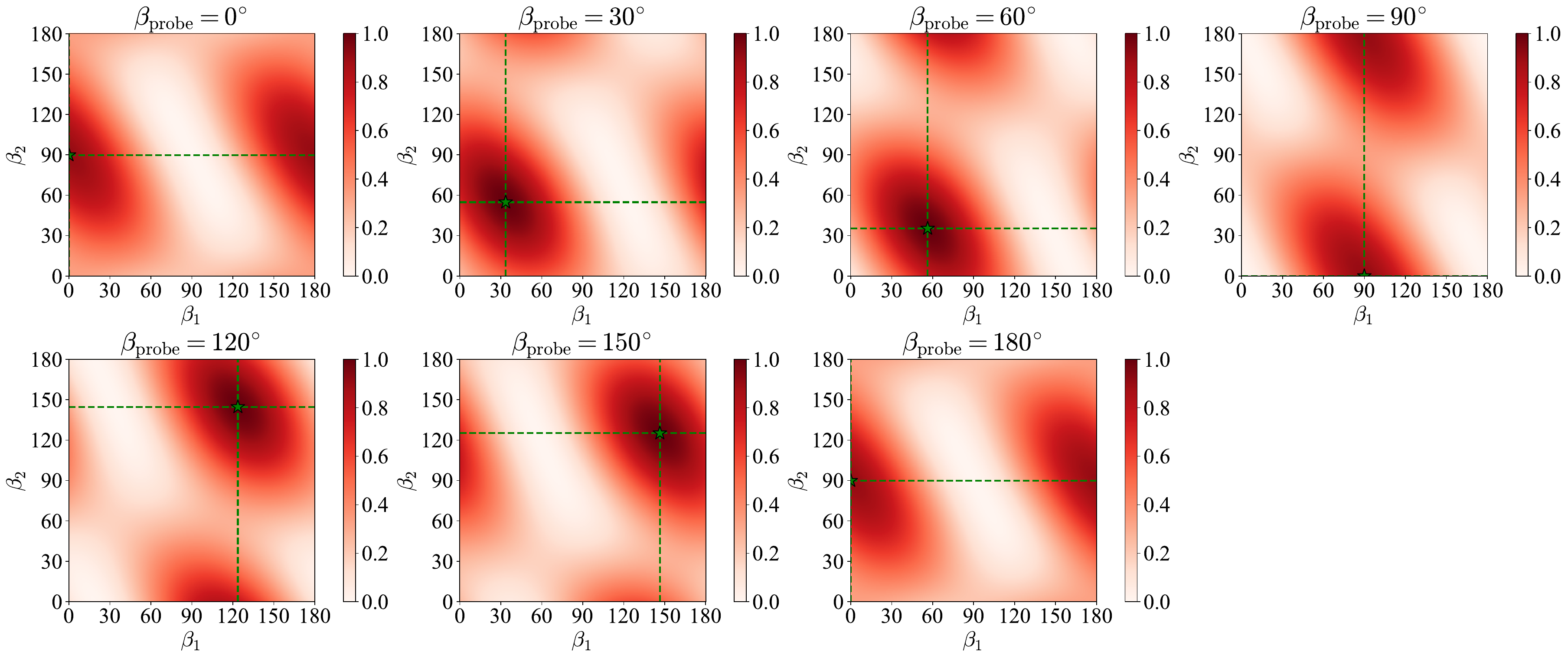}
    \caption{Polarization prefactor (from Eq. \ref{eq:pol-prefactor}) dependence on the polarization of individual beams ($\beta_1, \beta_2, \beta_{\rm probe}$) for a fixed collision angle $\theta_c = 60^{\circ}$. The maximal prefactor value was normalized to 1, with a normalizing factor of $\approx 32329$.}
    \label{appendix-fig:pol-60}
\end{figure}

Since the structure of this expression is not particularly transparent, we plot it numerically for different polarizations $(\beta_1, \beta_2, \beta_{\rm probe})$ for $\theta_c = 45^{\circ}$ (Fig. \ref{appendix-fig:pol-45}) and $\theta_c = 60^{\circ}$ (Fig. \ref{appendix-fig:pol-60}). The figures indicate that for each fixed $\beta_{\rm probe}$, there is an optimal pair of $(\beta_1, \beta_2)$ achieving the highest amplitude, which seems to be independent of $\beta_{\rm probe}$. Additionally, the optimal choice of $(\beta_1, \beta_2, \beta_{\rm probe})$ seems to be independent from $\theta_c$ as well. Changing $\theta_c$ only changes the overall scaling with $\theta_c \rightarrow 0^{\circ}$ leading to a larger amplitude.

\begin{figure}
    \centering
    \includegraphics[width=0.8\linewidth]{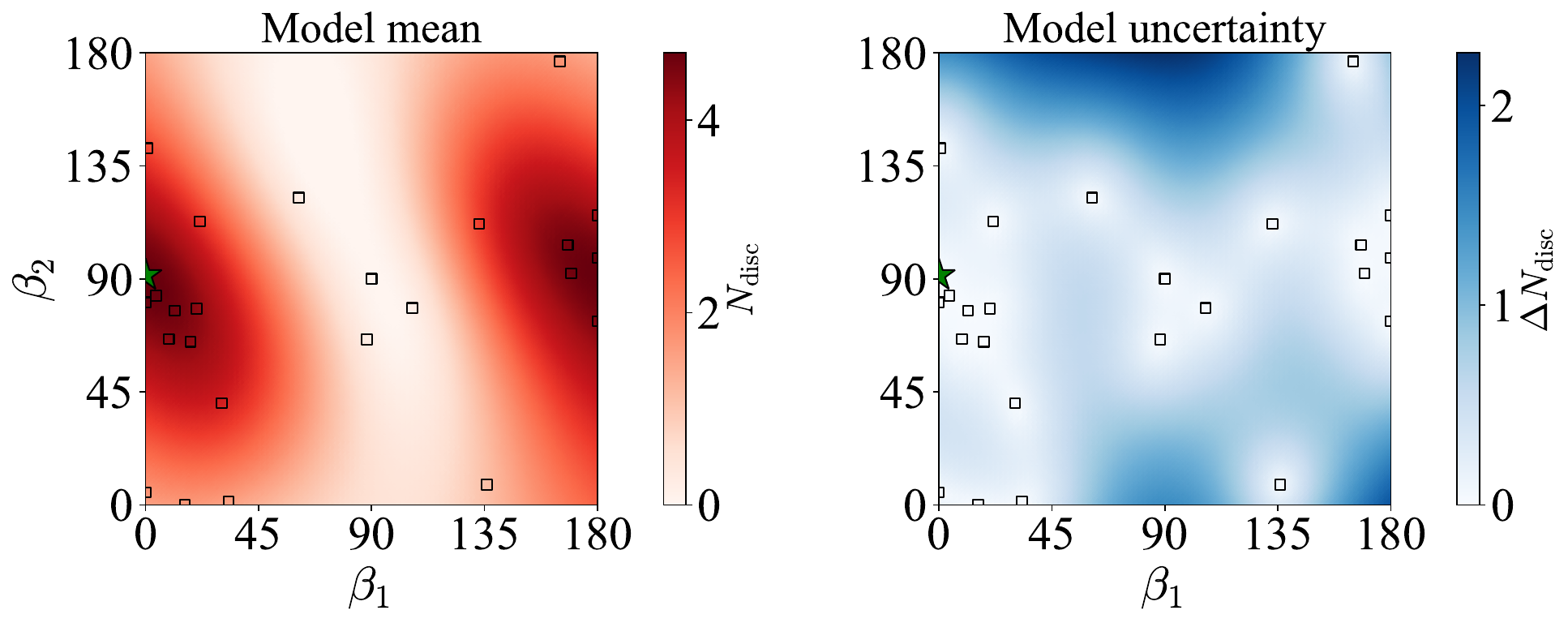}
    \caption{Optimization results in 2-dimensional parameter space: the linear polarization angle of pump beams $(\beta_1, \beta_2)$. Colormaps correspond to (left) the mean and (right) the 95\% confidence interval of the fitted Gaussian Process surrogate model. The optimization objective was to maximize the total number of discernible photons. The polarization of the probe is fixed $\beta_{\rm probe} = 0^{\circ}$ and the collision angle $\theta_c = 60^{\circ}$. Other parameters are similar to Fig. \ref{fig:3-beams}. Found optimum is $(\beta_1^*, \beta_2^*) \approx (0^{\circ}, 90^{\circ})$ achieving $ N_{\rm{disc}} \approx 4.7$. Compare the mean plot to the $\beta_{\rm probe}=0^{\circ}$ plots in Figs. \ref{appendix-fig:pol-45} and \ref{appendix-fig:pol-60}.}
    \label{appendix-fig:3-beam-pol-optim}
\end{figure}

The analytic findings agree with the optimization results in Fig. \ref{appendix-fig:3-beam-pol-optim}.

\end{document}